\documentclass[letterpaper,pdftex,twocolumn,amsmath,amssymb,groupedaddress,floatfix,superscriptaddress]{revtex4}

\usepackage{mathtools}

\DeclarePairedDelimiter{\evdel}{\langle}{\rangle}
\newcommand{\ev}{\operatorname{}\evdel}

\usepackage[scaled]{helvet}
\usepackage[utf8]{inputenc}
\usepackage[T1]{fontenc}
\usepackage{ae}
\usepackage{aecompl}
\usepackage{color}
\usepackage{textcomp}
\usepackage{makecell}
\usepackage{mathptmx}
\usepackage{float}
\usepackage{setspace}
\usepackage[a]{esvect}
\usepackage{pdfpages}
\usepackage{placeins}

\usepackage{appendix} 

\usepackage{footnote}

\usepackage{caption}
\DeclareCaptionLabelSeparator{bar}{ \rule[-0.2em]{0.3ex}{1em} }
\usepackage[sf,bf,tiny]{titlesec}
\titlespacing*{\section}{0pt}{1em}{0em}
\titlespacing*{\subsection}{0pt}{1em}{0em}

\definecolor{darkgray}{rgb}{0.25,0.25,0.25}
\definecolor{darkred}{rgb}{0.89,0.10,0.11}
\definecolor{darkblue}{rgb}{0.12,0.39,0.62}
\usepackage{url}

\usepackage[pdftex,breaklinks=true,colorlinks=true,citecolor=black,linkcolor=black,menucolor=black,urlcolor=darkblue,pdfborder={1 0 0}]{hyperref}

\usepackage{fixme} 

\usepackage{booktabs}

\begin{document}


	
\renewcommand{\figurename}{Figure}
\renewcommand{\thefigure}{\arabic{figure}}
\renewcommand{\tablename}{Table}
\renewcommand{\thetable}{\arabic{table}}
\renewcommand{\refname}{\large References}

\addtolength{\textheight}{1cm}
\addtolength{\textwidth}{1cm}
\addtolength{\hoffset}{-0.5cm}

\setlength{\belowcaptionskip}{1ex}
\setlength{\textfloatsep}{2ex}
\setlength{\dbltextfloatsep}{2ex}



\title{Visibility of minorities in social networks}



\author{Fariba Karimi}

\affiliation{GESIS -- Leibniz-Institute for the Social Sciences}
\affiliation{University of Koblenz-Landau}

\author{Mathieu G{\'e}nois}
\affiliation{GESIS -- Leibniz-Institute for the Social Sciences}

\author{Claudia Wagner}
\affiliation{GESIS -- Leibniz-Institute for the Social Sciences}
\affiliation{University of Koblenz-Landau}

\author{Philipp Singer}
\affiliation{GESIS -- Leibniz-Institute for the Social Sciences}
\affiliation{University of Koblenz-Landau}

\author{Markus Strohmaier}
\affiliation{GESIS -- Leibniz-Institute for the Social Sciences}
\affiliation{University of Koblenz-Landau}

\begin{abstract}
\begin{minipage}{0.70\textwidth}


Homophily can put minority groups at a disadvantage by restricting their ability to establish links with people from a majority group. This can limit the overall visibility of minorities in the network. Building on a Barab\'{a}si-Albert model variation with groups and homophily, we show how the visibility of minority groups in social networks is a function of (i) their relative group size and (ii) the presence or absence of homophilic behavior. We provide an analytical solution for this problem and demonstrate the existence of asymmetric behavior. Finally, we study the visibility of minority groups in examples of real-world social networks: sexual contacts, scientific collaboration, and scientific citation. Our work presents a foundation for assessing the visibility of minority groups in social networks in which homophilic or heterophilic behaviour is present. 

\end{minipage}
\end{abstract}
\maketitle


\noindent

\noindent Social networks are comprised of individuals with a variety of attributes, such as race, age, educational background, and gender. Commonly, these attributes are distributed unequally in the population. For example, in many schools across the United States and Europe, Asian or Blacks belong to a minority group \cite{moody2001race}, or women in science and engineering fields are a minority \cite{beede2011women}. In addition, homophily, the tendency to associate with similar others, is observed in many social networks, ranging from friendship to marriage to business partnerships \cite{mcpherson2001birds,moody2001race,baerveldt2004students,mislove2010you,fiore2005homophily}. One study has shown that in school friendships, Asians and Blacks are biased towards interacting with their own race at a rate >7 times higher than Whites and that homophily has a nonlinear relation with respect to relative group sizes \cite{currarini2010identifying}. 
However, the extent to which homophilic behaviour combined with group size differences can put minority groups at a disadvantage by limiting their visibility in social networks is not well understood.

Understanding the factors that impact the visibility of minorities has gained importance in recent years since algorithms have been widely used for ranking individuals in various application domains, including search engines \cite{brin1998anatomy,kleinberg1999hubs,horowitz2010anatomy}, recommender systems \cite{zhou2012state,king2010introduction}, or hiring processes \cite{boyd2014networked,miller2015can,chalfin2016productivity}. These rankings are critical, since they can influence the visibility of individuals and the opportunities afforded to them. Rankings are commonly based on the topological structure of networks, and hence, the position of individuals in their social network significantly influences their visibility. In particular, in networks in which one group of individuals is smaller in size (minority), visibility can have a crucial impact on the representation of the whole group. 
This raises fundamental questions about the effects of group sizes and the different mechanisms of tie formation on the visibility of minorities in social networks. 

\begin{figure*}[]
\centering
\includegraphics[width=.99\linewidth]{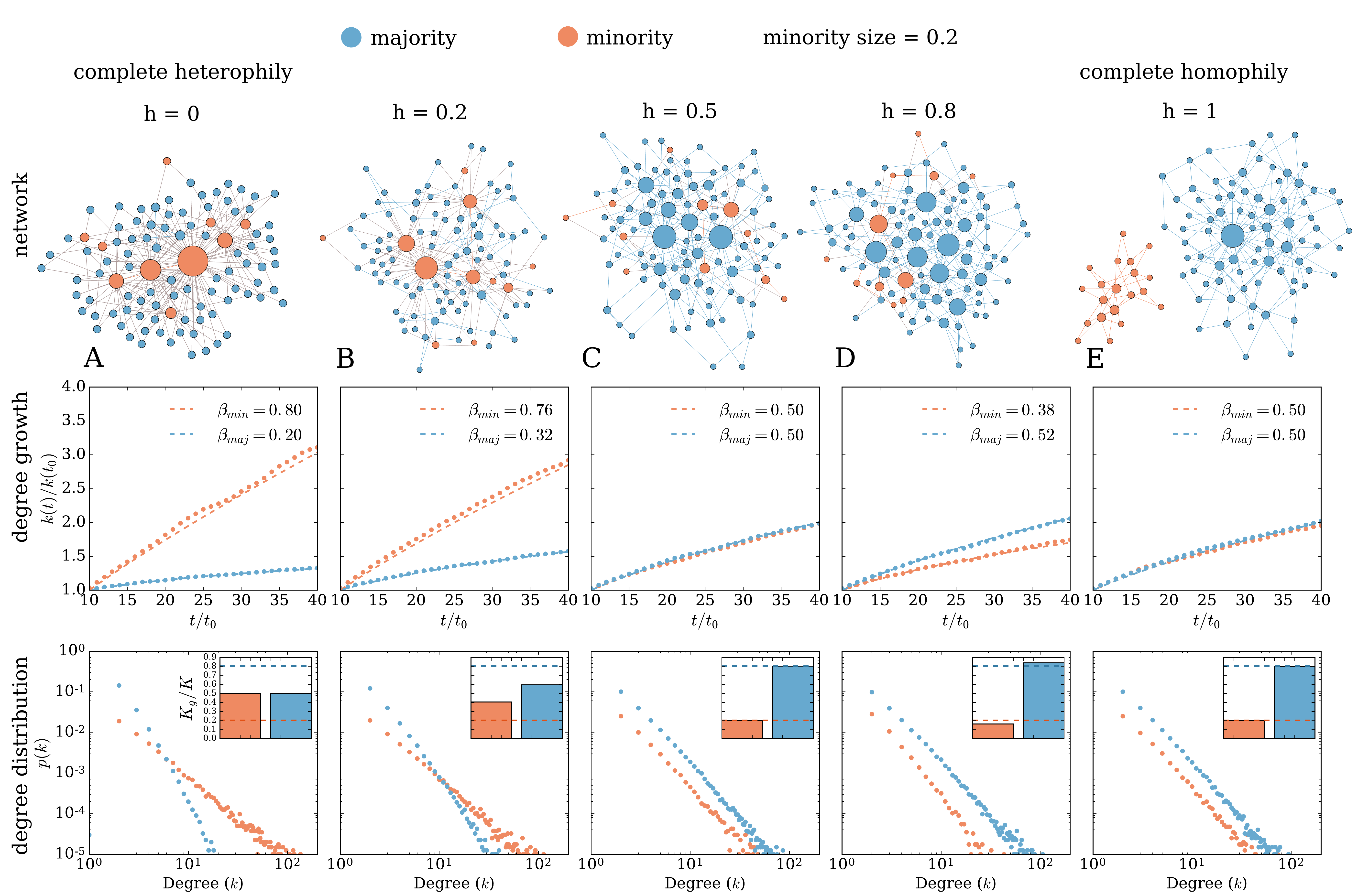}
\caption{\textbf{Disproportionate visibility and asymmetric effects of homophily on Barab\'{a}si-Albert networks with minority and majority groups.} The minority group (orange nodes) represents 20\% of the population. 
Homophily is regulated by parameter $h$.
Panel A represents a maximally heterophilic network ($h=0$). As homophily increases, nodes prefer to connect with nodes of the same color. Panel E
represents a maximally homophilic network ($h=1.0$).  The top row is a schematic of the network topology generated from the model (Eq.~\ref{eq:homophilic_BA}) for a small network with 100 nodes. The second row represents the resulting degree growth over simulation time steps and the third row represents the degree distribution generated from the model for two types of nodes. The inset in the third row depicts the share of total degree for minority and majority groups and the dashed lines show the fraction size of the group. In the heterophilic regime ($ 0 \leq h <0.5$), the degree of the minority group grows faster than majority. In the homophilic regime ($0.5<h\leq 1$) the growth of the degree slows down for minorities.  
The network size for the second and third row are generated for $N = 5000$ nodes. The results are averaged over 20 simulations.  }

\label{fig:fig1}
\end{figure*}

In this study, we utilize two main mechanisms for tie formation, homophily \cite{mcpherson2001birds}, and preferential attachment \cite{Barabasi99}, to systematically study how relative size differences between groups in social networks, with various levels of homophily, impact the visibility of nodes.  
In recent years, models have been proposed that consider homophily \cite{bramoulle2012homophily,currarini2010identifying}, or a combination of homophily and preferential attachment in the tie formation process \cite{papadopoulos2012popularity,de2013scale,avin2015homophily}. We build on these models by systematically exploring the parameter range for homophily and group size differences to explain the emergent properties of networks and their impact on the visibility of minority and majority groups. We define \emph{visibility} as the  importance of the node in the network, which is commonly measured by the degree of connectivity. Our results (cf. Figure ~\ref{fig:fig1} top row for an illustration) show that the visibility of nodes in such settings is \emph{disproportionate}---i.e. visibility is not proportional to the size of the group and varies by homophily.
Since the formation of links in such networks is driven by preferential attachment and homophily, we find that majority nodes are more visible in homophilic networks than expected, whereas minority nodes are more visible in heterophilic networks. Surprisingly, visibility has an \emph{asymmetrical} and non-linear effect in both homophilic and heterophilic regimes. We provide an analytical solution that predicts the exponent of the degree distribution and demonstrate the presence of this asymmetric effect. We show evidence of a disproportionate visibility in three empirical networks (sexual contacts, scientific collaboration, and scientific citation) with different ranges of homophily and group size.

In the following sections, we show the analytical and numerical results of the effect of homophily and group sizes on the degree and visibility of nodes in social networks. We then discuss the impact of the parameters on the ranking of nodes that belong to different groups. Finally, we show that our model captures network properties, such as the degree distributions and ranks of the majority and minority in empirical social networks with different group sizes and different degrees of homophily.

\section{Results}

\textbf{Model} 

\noindent We use the well-known model of preferential attachment proposed by Barab\'{a}si and Albert \cite{Barabasi99}, and we incorporate  homophily as an additional parameter to the model \cite{de2013scale,avin2015homophily}. Thus, the mechanism of tie formation in our model is influenced by the interplay between preferential attachment, via the degree of nodes, and homophily, via node attributes. A more general version of this model, known as the fitness model, was first proposed by Bianconi and Barab\'{a}si  \cite{bianconi2001competition}. In this model, the probability of a connection is the product of the degree and fitness of the node. However, the fitness of a node is assumed to be constant regardless of the presence of other nodes. In our model, the fitness of a node also depends on the attributes of other nodes.

We model social networks with two groups of nodes, in which all the nodes from the same group behave similarly. Let us call the two groups $a$ and $b$. We define a tunable homophily parameter $h$ that regulates the tendency of individuals to connect with other individuals who belong to the same group. The homophily parameter ranges between 0 to 1, $h \in [0,1]$, where $0$ means that the nodes from one group are attracted only by nodes from the other group (heterophily), $1$ means that the nodes connect only with similar nodes (homophily), and $0.5$ indicates a homogeneous mixing with respect to group affiliation. The model consists of $N$ nodes and two attributes that initially are assigned to two groups with given sizes. We call $f_a$ the fraction of nodes that belong to group $a$, and $f_b = 1 - f_a$ the fraction of nodes that belong to group $b$. We shall refer to group $a$ as the minority and group $b$ as the majority, so that $f_b \ge f_a$. At each time step, a newly arriving node $j$ randomly attaches to $m$ pre-existing nodes by evaluating their degree and group membership. Multiple linkage between two nodes is not allowed. The probability of node $j$ to connect to node $i$ is given by:

\begin{equation}
\Pi_{i} = \frac{h_{ij} k_{i}}{\sum_{l} h_{lj} k_{l}}
\label{eq:homophilic_BA}
\end{equation}
where $k_{i}$ is the degree of node $i$ and $h_{ij}$ is the homophily between the two nodes.

In general, the homophily parameter defines the probability of within and across group connections. For example, in the case of two groups, we have two homophily parameters: $h_{aa}$ (probability of connection between members of group $a$), $h_{bb}$ (probability of connection between members of group $b$), and the probability between groups ($h_{ab}$ and $h_{ba}$) are complementary probabilities ($h_{ab} = 1- h_{aa}$, $h_{ba} = 1- h_{bb}$) . As a simplification, one can assume homophily is regulated by only one parameter $h$, considering that homophily is symmetric and complementary: $h_{aa} = h_{bb} = h$ and  $h_{ab} = h_{ba} = 1 - h$. In this paper, we first provide the results for the simple case of symmetric homophily and then discuss asymmetric homophily. 

\begin{figure*}[]
\centering
\includegraphics[width=0.9\linewidth]{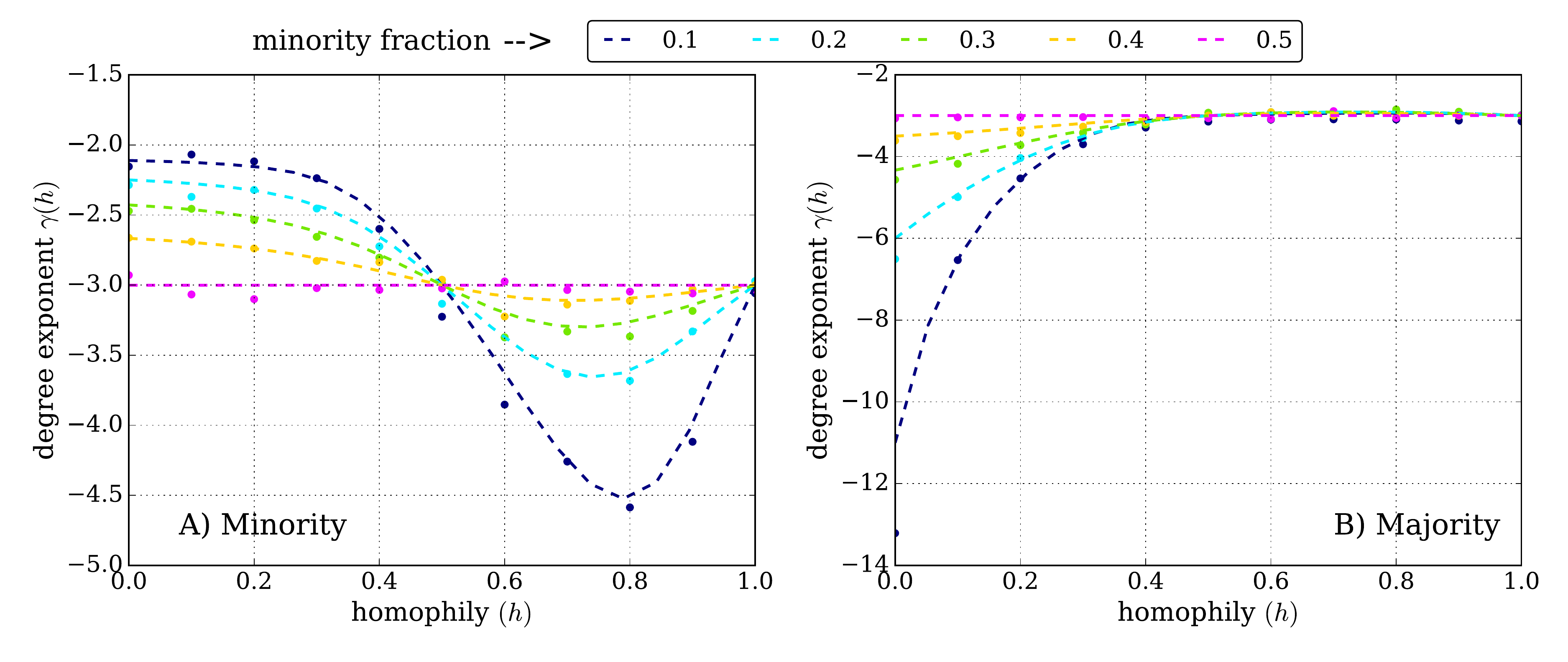}
\caption{\textbf{The analytical and numerical degree exponent of minority (A) and majority (B) versus homophily for various minority sizes.} The degree distribution follows a power-law function, $p(k) \propto k^{\gamma}$ in which the exponent of the distribution ($\gamma$) is shown by tuning homophily ($h$) and group sizes (shown by different colors). The dashed lines are the expected degree exponents derived from our analytical derivation (see Methods) and the dots represent the fitted value from the simulations of over 5,000 nodes. The analytical results are in excellent agreement with simulation. The size of the minority fraction is shown for the ranges between $0.1$ and $0.5$. For minority nodes (A), in the heterophily regime ($h < 0.5$), the degree exponent ranges from $-2$ to $-3$, which represents the advantage of these nodes to grow their degree to large values. In the homophilic regime ($h > 0.5$), the exponent shows a non-linear behaviour; as the degree exponent decreases, the advantage of the minorities to grow their degree becomes limited. However, this effect can be compensated in high homophilic regime by in-group support, which explains why the exponent increases for $h > 0.8$. For majorities (B), the heterophilic situation limits their advantage of growing their degree, in particular for small minority fractions. In homophilic regime, the exponent of the majority degree always remains close to $-3$ since the majority nodes do not gain extra advantage due to large group sizes. }
\label{fig:degree_dist_analytical}
\end{figure*}

\noindent \textbf{Degree growth}

\noindent Figure~\ref{fig:fig1} illustrates the dynamics of the degree growth by tuning homophily. The minority fraction is fixed to $0.2$. Our model is generalized and incorporates two types of network interactions. For the parameter range of $ 0 \leq h \leq 0.5$ the network is heterophilic, and for the range of $ 0.5 \leq h \leq 1$, the network is homophilic. In the heterophilic regime, the degree of the minority group grows faster than the degree of the majorities (see Fig.~\ref{fig:fig1} second row). The complete heterophilic case is equivalent to the formation of bipartite networks ($h=0$).  The difference in the degree growth reduces gradually as heterophily decreases, until we reach the homogeneous mixing case ($h=0.5$), in which groups do not matter anymore and we recover the original Barab\'{a}si-Albert growth model for both groups.

In the homophilic regime ($ 0.5 \leq h \leq 1$), the degree of the majority grows faster than the degree of the minority until a certain point $h = 0.8$. After that, the difference in growth decreases until we reach the fully homophilic case ($h = 1$) in which the network is split between the two groups, each having the same degree growth. The extreme homophilic case resembles societies in which women and men are completely segregated at schools or some universities, e.g., in Iran or Saudi Arabia \cite{mehran2003paradox}.

\begin{figure*}[]
\centering
\includegraphics[width=0.9\linewidth]{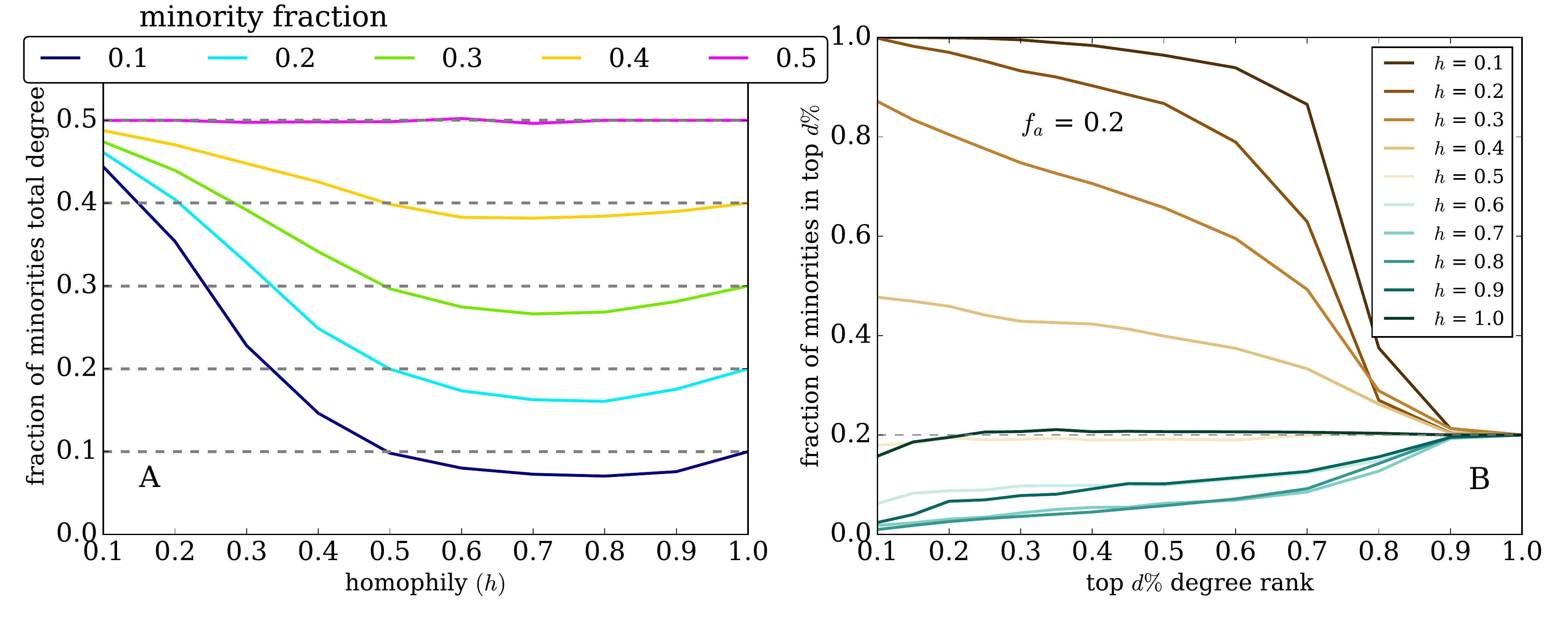}
\caption{\textbf{Visibility of minorities in relation to relative group size and homophily.} 
A) Average cumulative degree of the minority as a function of homophily, for different minority proportion (10\% - 50\%).
In a balanced population (0.5, pink line), both groups share half of the links, independently of the level of homophily. As the size of the minority decreases, the inequality in the share of degree increases.
In a homogeneous-mixing case ($h = 0.5$), the rank corresponds to the expected population size shown by the gray dashed lines. 
In heterophilic regimes ($ 0 \leq h < 0.5$), the minority takes advantage of the population size effect. In homophilic regimes ($ 0.5 < h \leq 1$), we observe that the degree of minorities is below the expectation and it is recovered only in the extreme homophilic case ($h = 1$) by full in-group support. 
B) Fraction of minority nodes that are found in the top d\% of nodes with the highest degree. The fraction of nodes belonging to the minority ($f_a$) is set to 0.2. If the group membership does not impact the attractiveness of nodes, we expect that the presence of the minority in the top d\% is proportional to its relative size (dashed line). However, the results are sensitive to the homophily parameter. In the heterophilic case ($0 \leq h < 0.5$), minorities are over-represented in the top d\%. In the homophilic case ($0.5<h \leq 1$), minorities are under-represented in the top d\%. In the case of homogeneous mixing ($h=0.5$) or complete homophily ($h = 1.0$), minorities are presented in the top d\% as expected from their relative size. 
}

\label{fig:degree_rank}
\end{figure*}


\noindent \textbf{Impact of homophily and group size on degree distribution and visibility}

\noindent Figure~\ref{fig:degree_dist_analytical} shows the exponent of the degree distribution for the minority (Fig.~\ref{fig:degree_dist_analytical}A) and majority (Fig.~\ref{fig:degree_dist_analytical}B) by tuning homophily and group sizes. We determine analytically the exact exponent of the degree growth and the degree distribution as a function of homophily ($h$) and minority size ($f_a$) (see Methods). The degree exponent illustrates the ability of nodes to stretch their degrees to high values and thus receive more visibility. Let us denote the degree distribution $p(k) \sim k^{\gamma}$, where $\gamma$ is the exponent of the degree distribution. When both groups are of equal size ($f_a = 0.5$), the model recovers the exponent $\gamma = -3$ for the degree distributions of both groups, as predicted from the classical  Barab\'{a}si-Albert model. In the heterophilic regime ($h<0.5$), as the size of the minority decreases, the exponent of the degree distribution of the minority increases, which indicates that the distribution stretches to larger values. The opposite situation occurs for majorities; as the size of the minority decreases, the exponent of the degree decreases which indicates that majorities are limited in stretching their degree to large values.

The homophilic regime ($h>0.5$) exhibits interesting behaviour. While the exponent of the degree distribution for the majority does not change much when we tune group size or homophily, there is a non-linear effect for the minority. As homophily increases, the exponent decreases until we reach a certain homophily value ($h \simeq 0.8$), and increases afterwards (see Fig.~\ref{fig:degree_dist_analytical}A). In the extreme homophilic case ($h = 1.0$) the degree growth of both groups is similar to the homogeneous mixing case ($h = 0.5$) and so are the exponents of the degree distributions. 

This non-linear behaviour can be explained by the interplay between homophily and relative group size differences. Both determine the amount of competition faced by the nodes of different groups. For the majority, heterophilic conditions are not beneficial, since nodes are mostly attracted by the minority, which as a consequence becomes extremely popular. Therefore, majority nodes have difficulties competing for the attention of the newly arriving nodes. In the homophilic regime, the majority is relatively indifferent because they compete for attention mostly among themselves. 

For the minority, heterophilic situations are most beneficial. They receive the most attention from the majority, and the competition for attention among minority nodes is relatively low since they are a small group. In homophilic situations, it is much more difficult for minority nodes to attract newly arriving nodes due to the competition with the majority, which is not only larger in size but also contains more popular nodes. However, in the case of extreme homophily, no competition exists between the nodes of different groups, and thus both groups compete only among themselves. The degrees of nodes in both groups grow similarly and their degree distributions are the same as in the homogeneously mixed case with the only difference that the network is split between the groups.

\noindent \textbf{Visibility of minorities in top ranks}

\noindent So far we have observed that homophily and the differences in group sizes have an effect on the degree growth and the degree distribution of groups. Although these findings may be rather trivial, the outcome of such interactions on the visibility of groups is striking.

Figure \ref{fig:degree_rank}A depicts the average total degree share of the minority as a function of homophily. Colors represent different minority sizes. The results for the majority group are complementary. In the extreme heterophilic case ($h = 0$), a minority group that represents 20\% of the total population (light blue line) receives more than 40\% of all degrees. This result resembles the idea of majority illusion in which the majority of nodes perceives the opinion of the minority as the majority opinion because they are exposed mainly to minority nodes \cite{lerman2015majority}. As the homophily between groups increases up to 0.5, the average total degree decreases to what we would expect from the size of the minorities (dashed gray lines). In the homophilic case ($0.5 \leq h \leq 1$), the degree drops below what we would expect from the population size, and thus, the minority group as a whole is penalized for the homophilic behaviour. In the extreme homophilic situation ($h = 1$), the minority group can take advantage of full in-group support and as a result the degree returns to the expectation that is proportional to the group size.

If we wish to examine only the top-ranked nodes, which is a realistic scenario for users who want to explore a ranked list of items (as in search user interfaces), the results are even more striking. Figure \ref{fig:degree_rank}B illustrates the probability of finding minorities in the top $d$\% of nodes ranked by degree. For example $d = 0.2$ means the fraction of nodes in the top 20\% of the nodes ranked by degree. 
In the heterophilic case (brown shades), nodes from the minority are overrepresented in the top ranked nodes. In the homophilic case (green shades), nodes from the minority are underrepresented, an effect especially important for small top $d$\,\%. Given the fact that nodes with high degree are very stable in terms of their rank \cite{ghoshal2011ranking}, these results suggest that in homophilic networks, the majority stabilizes its position at high ranks and leaves little opportunity for minorities to appear in the top ranks. In heterophilic cases, the roles are reversed: minority nodes stabilize their position at high ranks.    Given the fact that many social networks are homophilic with respect to attributes such as gender or ethnicity, our results suggest that in homophilic networks majorities occupy the high ranks and minorities tend to appear towards the lower ranks compared to what we would expect from the minority size.

\begin{figure*}[]
\centering

\includegraphics[width=1\linewidth]{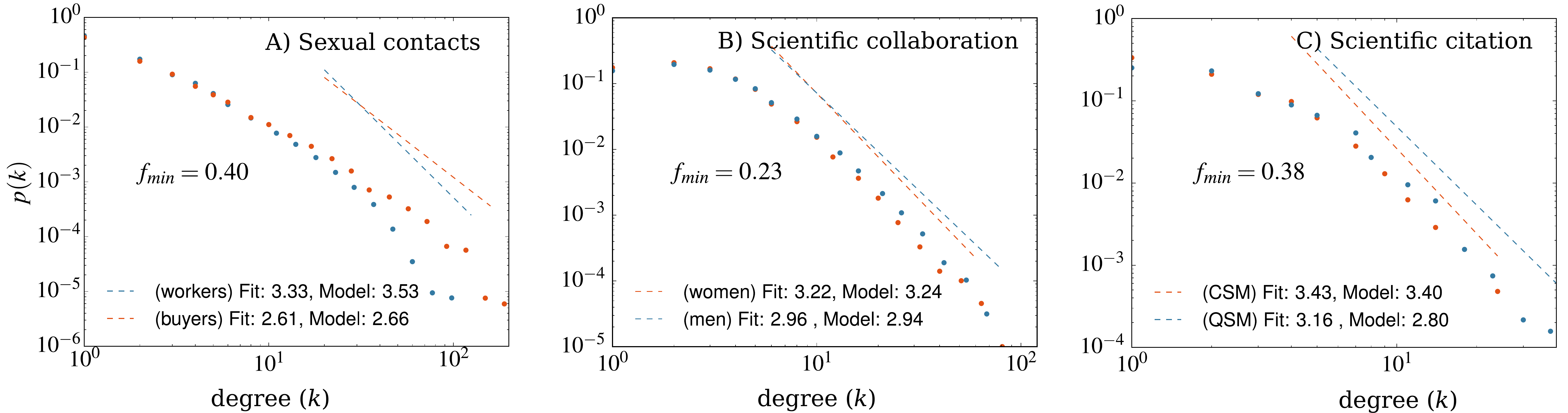}

\caption{\textbf{Degree distribution of majority and minority groups in three empirical networks.} A) Sexual contact network with sex-workers (blue) and sex-buyers (orange). B) Collaboration network with men (blue) and women (orange). C) American Physical Society (APS) citation network among two category of topics: Classical Statistical Mechanics (CSM, orange) and Quantum Statistical Mechanics (QSM, blue). The fraction of minorities are shown in the plots ($f_{min}$). The dashed line is the fitted line using the maximum likelihood estimate. The exponent of the fit (Fit) is compared with the analytical exponent derived from our model (Model). Our model is able to produce a realistic degree exponent for empirical networks with various types of homophily and group sizes.}
\label{fig:deg_dist_empirical}
\end{figure*}

\begin{figure*}[]
\centering
\includegraphics[width=1\linewidth]{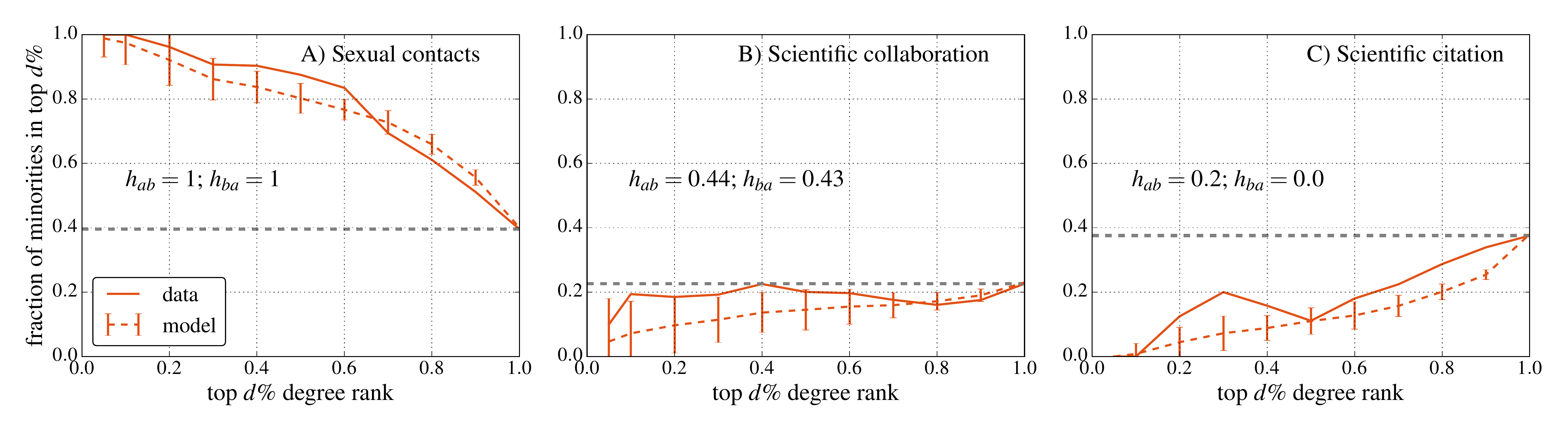}
\caption{\textbf{Visibility of minority groups in the top d\% degree rank for three empirical networks}. A) Sexual contact network (minority = sex buyers). B) Scientific collaboration network (minority = women). C) Scientific citation network (minority = Classical Statistical mechanics (CSM)). The solid line is measured from the empirical network and the dashed orange lines are the predicted values from the synthetic networks with a similar homophily parameter for $5,000$ nodes and averaged over $100$ simulations. The dashed gray line is the relative size of the minority. In the heterophilic case of the network of sexual contacts, the minority is overrepresented with respect to its size. In the collaboration network where homophily is moderate, the minority is underrepresented or it is  close to its relative size. In the case of the citation network that is extremely homophilic, the minority is highly underrepresented. These results provide empirical evidence for a visibility bias in empirical networks.}
\label{fig:topk_empirical}
\end{figure*}

\noindent \textbf{Visibility bias in empirical social networks}

\noindent We provide evidence for the emergence of degree inequality and rank differences in real social networks via three collected empirical social networks that exhibit various ranges of group size and homophily: sexual contacts, scientific collaboration and scientific citation.

We first determine the value of the homophily parameter in empirical networks. Established methods to quantify homophily include assortativity mixing \cite{newman2003mixing} and dyadicity \cite{park2007distribution}. These measures are to quantify the significance level of outgroup links compared to random expectation. However, real social networks do not necessarily exhibit symmetric homophilic behaviours. Observing only the edges between groups does not capture this potential asymmetric behaviour between groups. For example, if homophily among minorities (fraction size = $0.2$), is $0.1$ and for majorities is $0.7$, the assortativity by definition will be close to zero and similar to the case in which homophily is equal to $0.5$ for both groups. However, in this case we would expect that the number of edges that exist within the majority group is far greater than the number of edges within the minority group, after correcting for the group size. Therefore, to fully grasp asymmetric homophilic behavior, we need to consider the fraction of links that run between groups and within groups. Given the number of links that run between each group and the relation between group sizes, homophily, and degree exponent, we can analytically determine the homophily parameter for each group (see Methods and Fig.~\ref{fig:exponents_asym}).

The analytical derivation enables us to accurately estimate the value of the homophily parameter in empirical networks by using only the number of edges within each group given our model parameters. We then focus on three examples of networks that exhibit high heterophily (sexual contacts), moderate homophily (scientific collaboration) and high homophily (scientific citation). We assume that all networks are undirected and we focus on one node attribute (e.g. gender or scientific field).


The first network captures sexual contacts between sex-workers and sex-buyers  \cite{rocha2011simulated}. The network consists of 16,730 nodes and 39,044 edges. There are 10,106 sex-workers and  6,624 sex-buyers (minority size $f_a = 0.4$). In this network, no edges among members of the same type exist and consequently the homophily parameter is equal to 0 for both groups, $h_{aa} = h_{bb} = 0$. 



The second network, which exhibits moderate homophily and relative group size difference, depicts scientific collaborations in computer science extracted from DBLP \cite{DBLP}. We used a new method that combines names and images to infer the gender of the scientists with high accuracy \cite{karimi2016inferring}. We focus on a 4-years snapshot of the network. After removing ambiguous names, the resulting network consists of 280,200 scientists and 750,601 edges (paper co-authorships) with 63,356 female scientists and 216,844 male scientists ($f_a = 0.23$).
We measure the homophily among women ($h_{aa} = 0.56$) and among men ($h_{bb} = 0.57$) and find a slight tendency for men to connect more among themselves than women.


The last network captures scientific citations in the American Physical Society (APS) corpus that exhibits strong homophily. Citation networks reveal how much attention communities around different topics attribute to each other. We use PACS identifier to select papers on the same topics. In this case we chose statistical physics, thermodynamics and nonlinear dynamical system sub-fields (PACS = 05). Within a specific sub-field there are many sub-topics that form communities of various sizes. To make the data comparable with our model, we choose two sub-topics that are relevant, namely classical statistical mechanics (CSM - 05.20.-y) and quantum statistical mechanics (QSM - 05.30.-d). 
The resulting network consist of 1,853 scientific papers and 3,627 citation links. The minority group in these two sub-topics is CSM ($f_a = 0.38$).
We find weaker homophily for the CSM papers ($h_{aa} = 0.8$) than for QSM papers ($h_{bb} = 1$), which indicates asymmetric, homophilic behavior in citation networks. 


To evaluate our model against the data, we compare the exponent of the empirical degree distribution with the exponent generated from our model given the same empirical homophily and group size values. To estimate the exponent of the empirical degree distribution we use the maximum-likelihood fitting method \cite{clauset2009power,alstott2014powerlaw}. The exponent of the degree distribution generated from the model is calculated analytically (see Methods).


Figure \ref{fig:deg_dist_empirical} displays degree distribution of minorities and majorities in the three empirical networks. The exponent of the degree distributions of networks generated with our model agrees well with the empirical degree exponents using maximum-likelihood fit.

Similar to the previous section, we also examine the top nodes ranked by degree. Figure \ref{fig:topk_empirical} illustrates the probability of finding minorities in the top $d$\,\% of nodes ranked by degree. In the heterophilic case of the network of sexual contacts Fig.~\ref{fig:deg_dist_empirical}A, the minority is overrepresented with respect to its size.  In the
scientific collaboration Fig.~\ref{fig:topk_empirical}B in which homophily is moderate, the minority rank is close to its relative size. In the case of the scientific citation Fig.~\ref{fig:topk_empirical}C which is extremely homophilic, the representation of the minority is highly underestimated. We provide the results of the ranks in synthetic networks with similar homophilic parameters (dashed orange lines). Despite the simplicity of the model compare to the empirical data, the majority of ranks fall well within the standard deviation of the model. These results provide empirical evidence for a visibility bias in empirical networks and the usefulness of the model to capture biases. 



\section{Discussion}

We demonstrate analytically and numerically that the visibility of nodes (measured by degree) is influenced by relative group size difference and homophily and the visibility has asymmetric and non-linear properties. As the size of a minority group decreases, minority can benefit more from heterophilic interactions and suffer from homophilic interactions. However, minorities can recover their visibility by full in-group support. Although our model makes simple assumptions such as all members of the same group behave similarly and are equally active, it lays a theoretical foundation for studying how the inherent properties of networks would lead to biases in visibility or ranking of groups, in particular minorities. 

Our work can be extended in multiple ways. First, new ranking algorithms can be devised to harness relative group size differences and homophily to ensure the representativeness of minorities and correct for potential biases. Second, the model can be extended to account for directionality and multiple attributes in networks and multiplex networks. Third, this model can be used to study community detections in annotated networks \cite{newman2016structure}, sampling hard-to-reach populations \cite{shaghaghi2011approaches} or evaluating the performance of classifiers in machine learning tasks \cite{hardt2016equality,dwork2012fairness}. We anticipate that this work will inspire more empirical and theoretical exploration on the impact of network structure on the visibility and ranking of minorities to help establish more equality and fairness in society.

\section{Methods}

Here, we provide the analytical derivation of degree growth and the exponent of the degree distribution of the model. We do this using two approaches; exact derivation and continuum approximation (see Appendix).

\subsection{Exact degree dynamics}
Let $K_a(t)$ and $K_b(t)$ be the sum of the degrees of nodes from group $a$ and $b$ respectively. Since the overall growth of the network follows a Barab\'asi-Albert process, the evolution of these quantities verify:
\begin{equation}\label{eq:kakb}
    K_a(t) + K_b(t) = K(t) = 2mt
\end{equation}
where $m$ is the number of new links in the network at each time step $t$. Let us denote the relative fraction of group size for each group as $f_a$ and $f_b$. The evolution of $K_a$ and $K_b$ is given in discrete time by:

\small
\begin{widetext}
\begin{equation}
 \left\{
  \begin{array}{l}
    K_a(t+\Delta t) = K_a(t) + m\left(f_a\left(1 + \dfrac{h_{aa}K_a(t)}{h_{aa}K_a(t) + h_{ab}K_b(t)}\right) + f_b\dfrac{h_{ba}K_a(t)}{h_{bb}K_b(t) + h_{ba}K_a(t)}\right)\Delta t\\
    \\
    K_b(t+\Delta t) = K_b(t) + m\left(f_b\left(1 + \dfrac{h_{bb}K_b(t)}{h_{bb}K_b(t) + h_{ba}K_a(t)}\right) + f_a\dfrac{h_{ab}K_b(t)}{h_{aa}K_a(t) + h_{ab}K_b(t)}\right)\Delta t\\
  \end{array}
 \right.
\end{equation}
\normalsize
which in the limit $\Delta t \rightarrow 0$ gives:
\small
\begin{equation}\label{eq:evol_ka}
\left\{
  \begin{array}{l}
    \dfrac{dK_a}{dt} = m\left(f_a\left(1 + \dfrac{h_{aa}K_a(t)}{h_{aa}K_a(t) + h_{ab}K_b(t)}\right) + f_b\dfrac{h_{ba}K_a(t)}{h_{bb}K_b(t) + h_{ba}K_a(t)}\right)\\
    \\
    \dfrac{dK_b}{dt} = m\left(f_b\left(1 + \dfrac{h_{bb}K_b(t)}{h_{bb}K_b(t) + h_{ba}K_a(t)}\right) + f_a\dfrac{h_{ab}K_b(t)}{h_{aa}K_a(t) + h_{ab}K_b(t)}\right)\\
  \end{array}
\right.
\end{equation}
\end{widetext}

\normalsize

\begin{figure}[]
\centering
\includegraphics[width=0.8\linewidth]{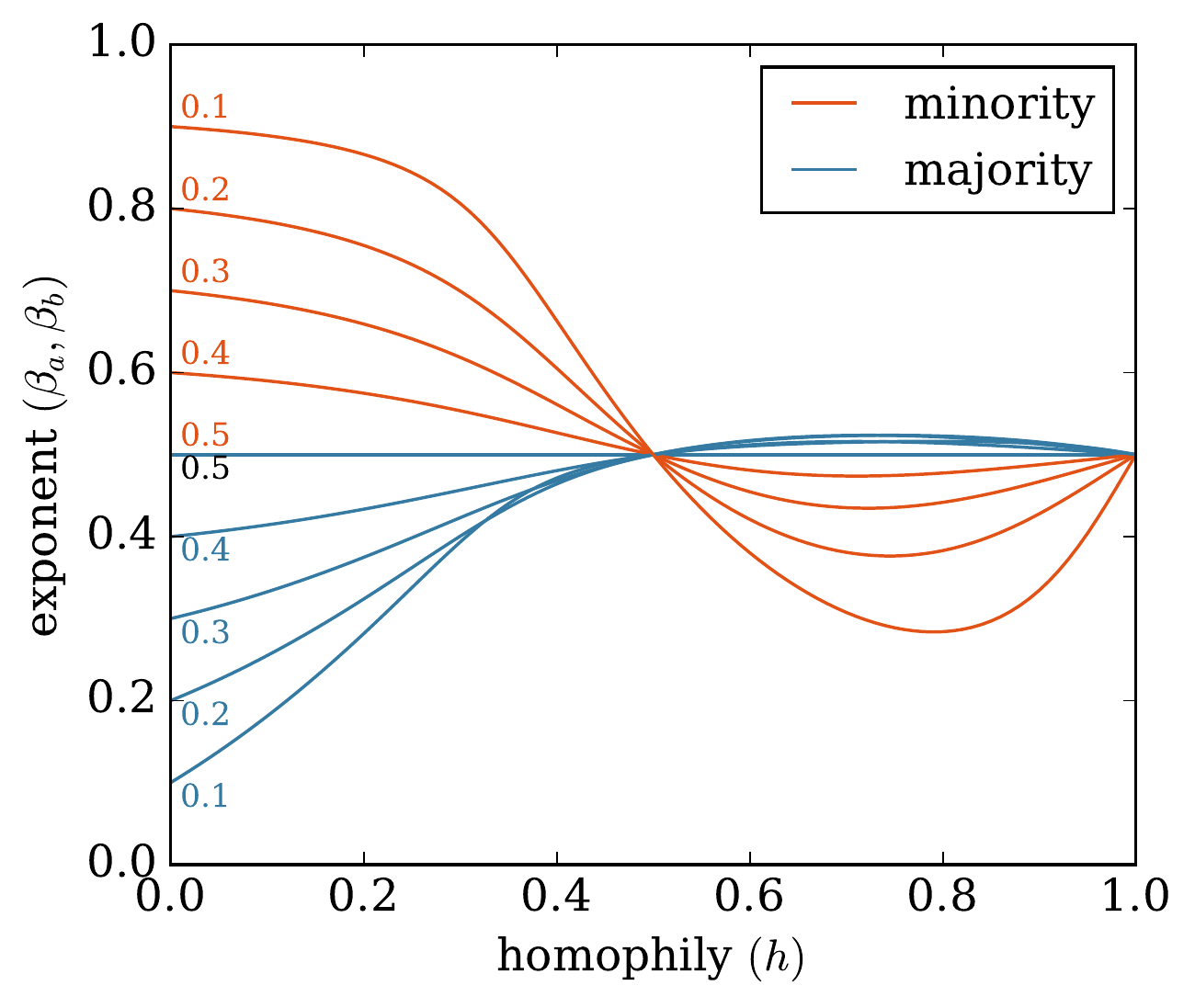}
\caption{\textbf{Evolution of the exponents for the degree growth, symmetrical homophily.} The exponents $\beta_a$ (minority) and $\beta_b$ (majority) are defined in eqs.~(\ref{eq:ca}) and (\ref{eq:cb}). $h = h_{aa} = h_{bb}$ is the homophily parameter and the numbers indicate the fraction of nodes belonging to the minority group (parameter $f_a$).}
\label{fig:exponents}
\end{figure}

These equations verify that for $h_{aa} = h_{bb} = 0$ and $h_{ab} = h_{ba} = 1$ (perfectly heterophilic network) we get:
\small
\begin{equation}
\left\{
  \begin{array}{l}
    \dfrac{dK_a}{dt} = m\\
    \\
    \dfrac{dK_b}{dt} = m\\
  \end{array}
\right.
\end{equation}
\normalsize
and thus for the evolution of the degree of a single node:
\small
\begin{equation}
\left\{
  \begin{array}{l}
    \dfrac{dk_a}{dt} = mf_b\dfrac{k_a}{\sum_i q_ik_i} = mf_b\dfrac{k_a}{K_b(t)} = f_b\dfrac{k_a}{t}\\
    \\
    \dfrac{dk_b}{dt} = mf_a\dfrac{k_b}{\sum_i q_ik_i} = mf_a\dfrac{k_b}{K_a(t)} = f_a\dfrac{k_b}{t}\\
  \end{array}
\right.
\end{equation}
\normalsize
which gives:
\begin{equation}
\left\{
  \begin{array}{l}
    k_a \propto t^{f_b}\\
    k_b \propto t^{f_a}\\
  \end{array}
\right.
\end{equation}
Similarly, for $h_{aa} = h_{bb} = 1$ and $h_{ab} = h_{ba} = 0$ (perfectly homophilic network) we get:
\small
\begin{equation}
\left\{
  \begin{array}{l}
    \dfrac{dK_a}{dt} = 2mf_a\\
    \\
    \dfrac{dK_b}{dt} = 2mf_b\\
  \end{array}
\right.
\end{equation}
\normalsize
and thus for the evolution of the degree of a single node:
\small
\begin{equation}
\left\{
  \begin{array}{l}
    \dfrac{dk_a}{dt} = mf_a\dfrac{k_a}{\sum_i q_ik_i} = mf_a\dfrac{k_a}{K_a(t)} = \dfrac{k_a}{2t}\\
    \\
    \dfrac{dk_b}{dt} = mf_b\dfrac{k_b}{\sum_i q_ik_i} = mf_b\dfrac{k_b}{K_b(t)} = \dfrac{k_b}{2t}\\
  \end{array}
\right.
\end{equation}
\normalsize
which gives:
\begin{equation}
\left\{
  \begin{array}{l}
    k_a \propto t^{1/2}\\
    k_b \propto t^{1/2}\\
  \end{array}
\right.
\end{equation}
\normalsize

Let's make the hypothesis that $K_a(t)$ and $K_b(t)$ are linear functions of time, so that $K_a(t) = Cmt$ and $K_b(t) = (2-C)mt$ given Eq.~(\ref{eq:kakb}). Using Eq.~(\ref{eq:evol_ka}), we thus have:
\small
\begin{widetext}
\begin{equation}
    \dfrac{dK_a}{dt} = Cm = m\left(f\left(1 + \dfrac{h_{aa}Cmt}{h_{aa}Cmt + h_{ab}(2mt - Cmt)}\right) + (1-f)\dfrac{h_{ba}Cmt}{h_{bb}(2mt - Cmt) + h_{ba}Cmt}\right)
\end{equation}
\normalsize
which can be rewritten as:
\begin{equation}\label{eq:C}
\begin{split}
  (h_{aa} - h_{ab})(h_{ba} - h_{bb})C^3 \\
  + ((2h_{bb} - (1-f)h_{ba})(h_{aa} - h_{ab}) + (2h_{ab} - f(2h_{aa} - h_{ab}))(h_{ba} - h_{bb}))C^2 \\
  + (2h_{bb}(2h_{ab} - f(2h_{aa} - h_{ab})) - 2fh_{ab}(h_{ba} - h_{bb}) - 2(1-f)h_{ba}h_{ab})C \\
  - 4fh_{ab}h_{bb} = 0
\end{split}
\end{equation}
\end{widetext}

\begin{figure*}[]
\centering
\includegraphics[width=0.23\linewidth]{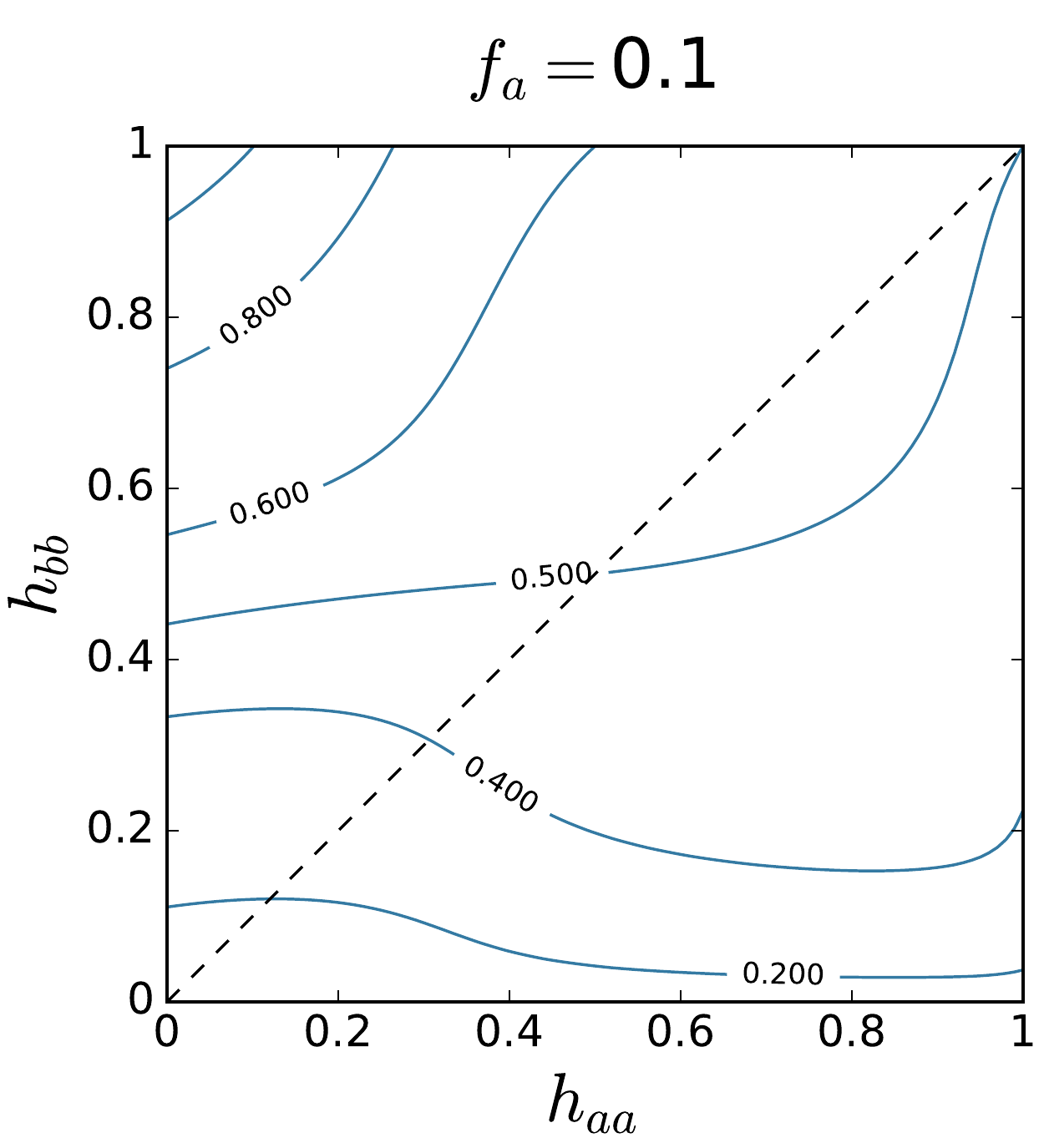}
\includegraphics[width=0.23\linewidth]{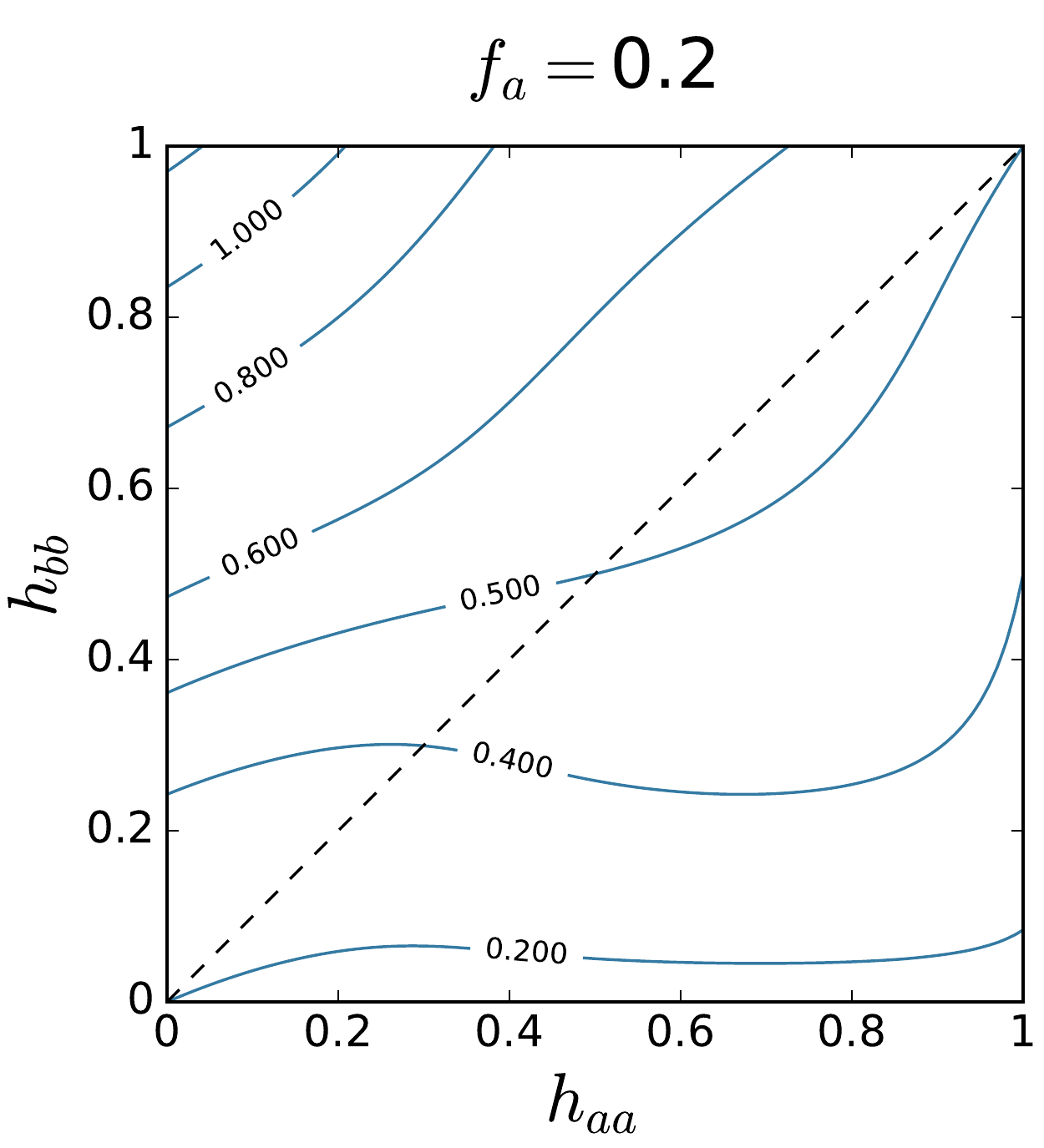}
\includegraphics[width=0.23\linewidth]{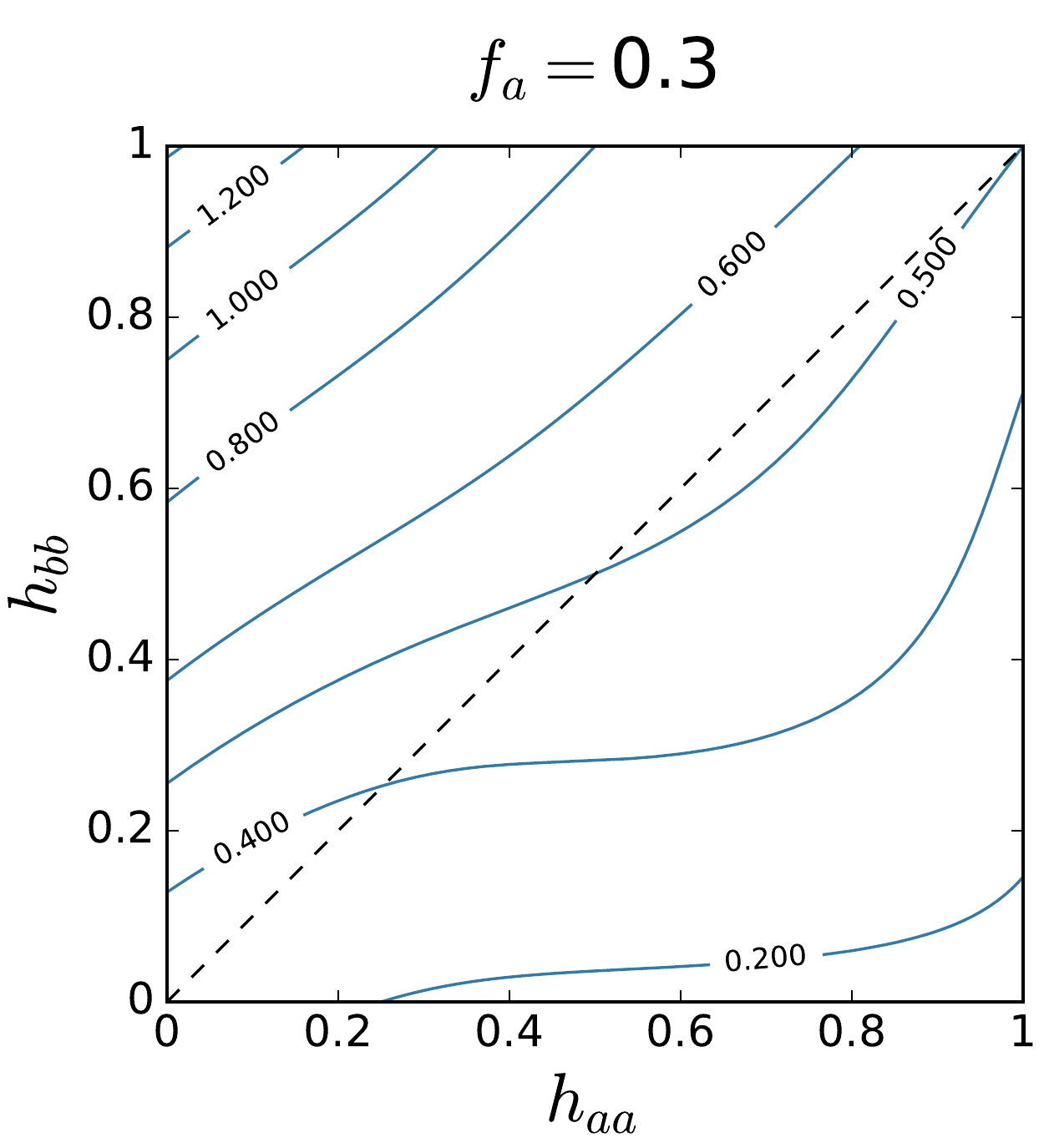}
\includegraphics[width=0.23\linewidth]{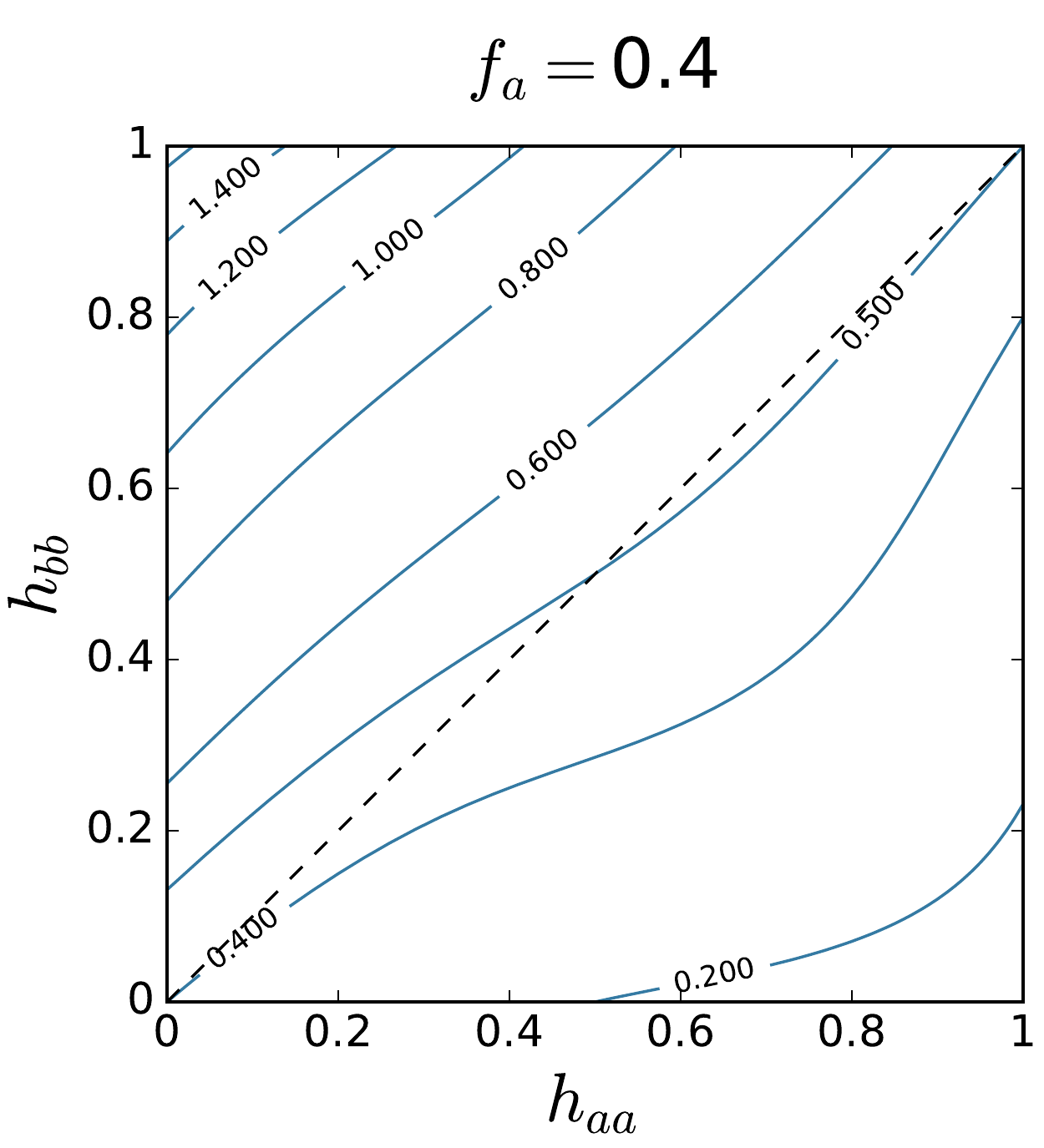}

\includegraphics[width=0.23\linewidth]{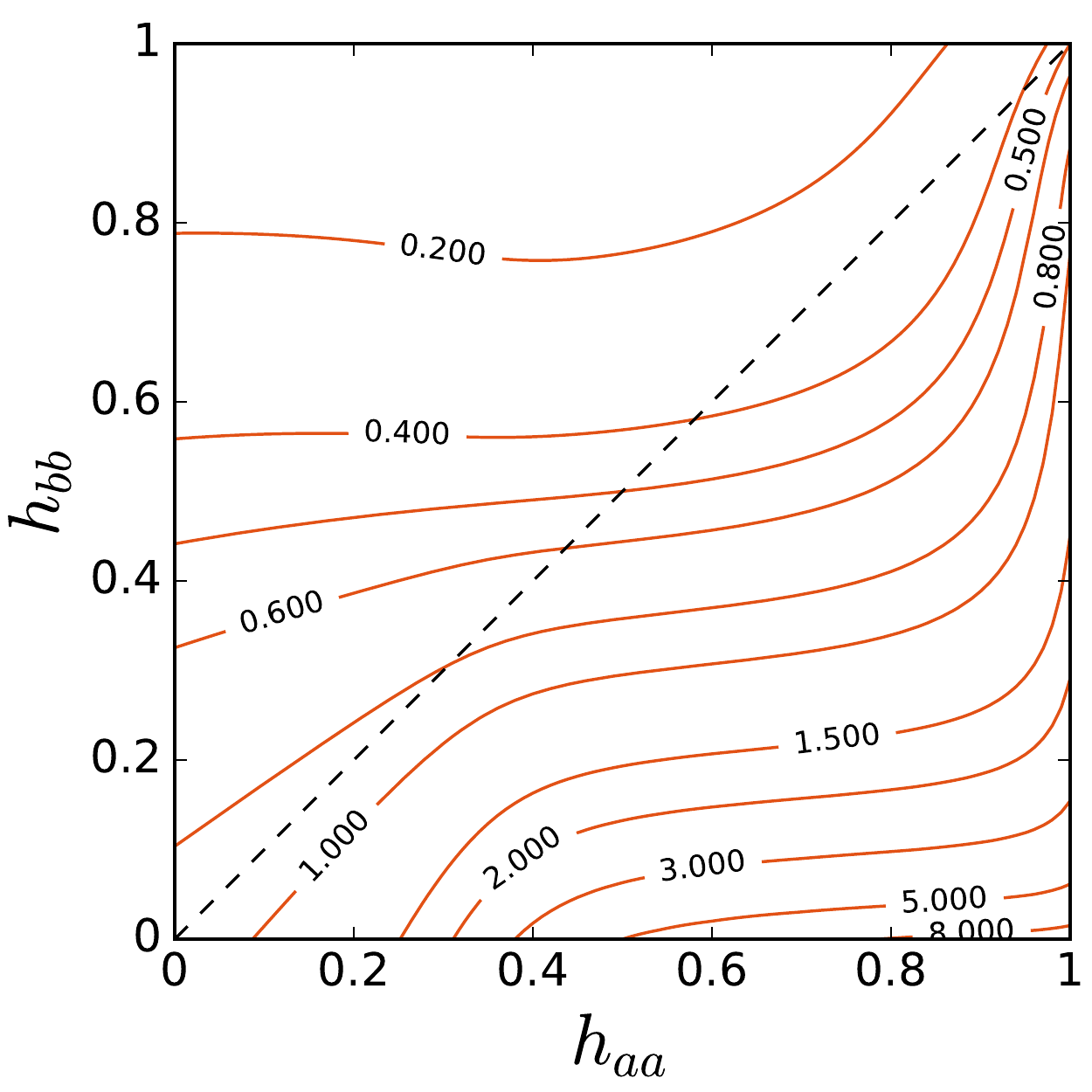}
\includegraphics[width=0.23\linewidth]{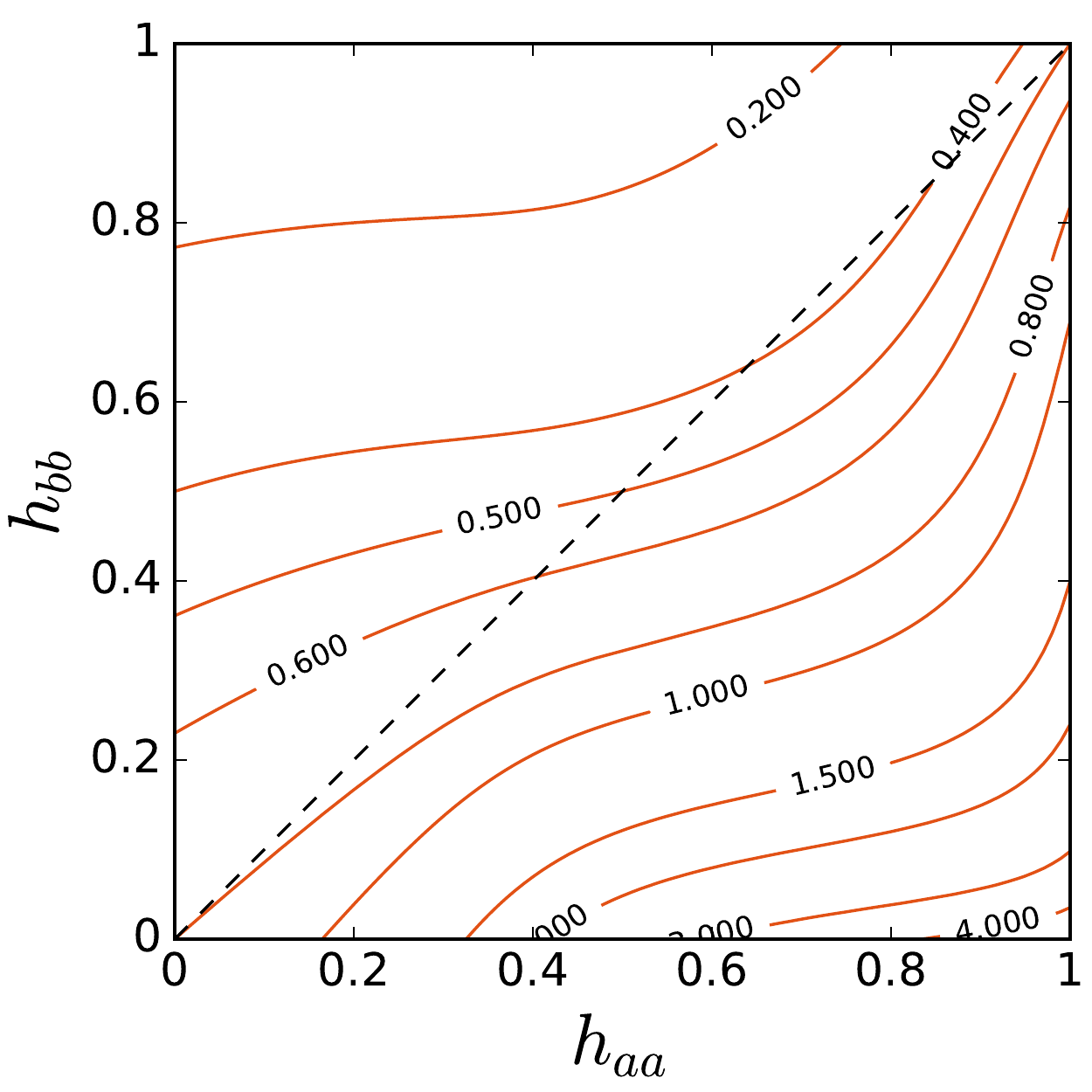}
\includegraphics[width=0.23\linewidth]{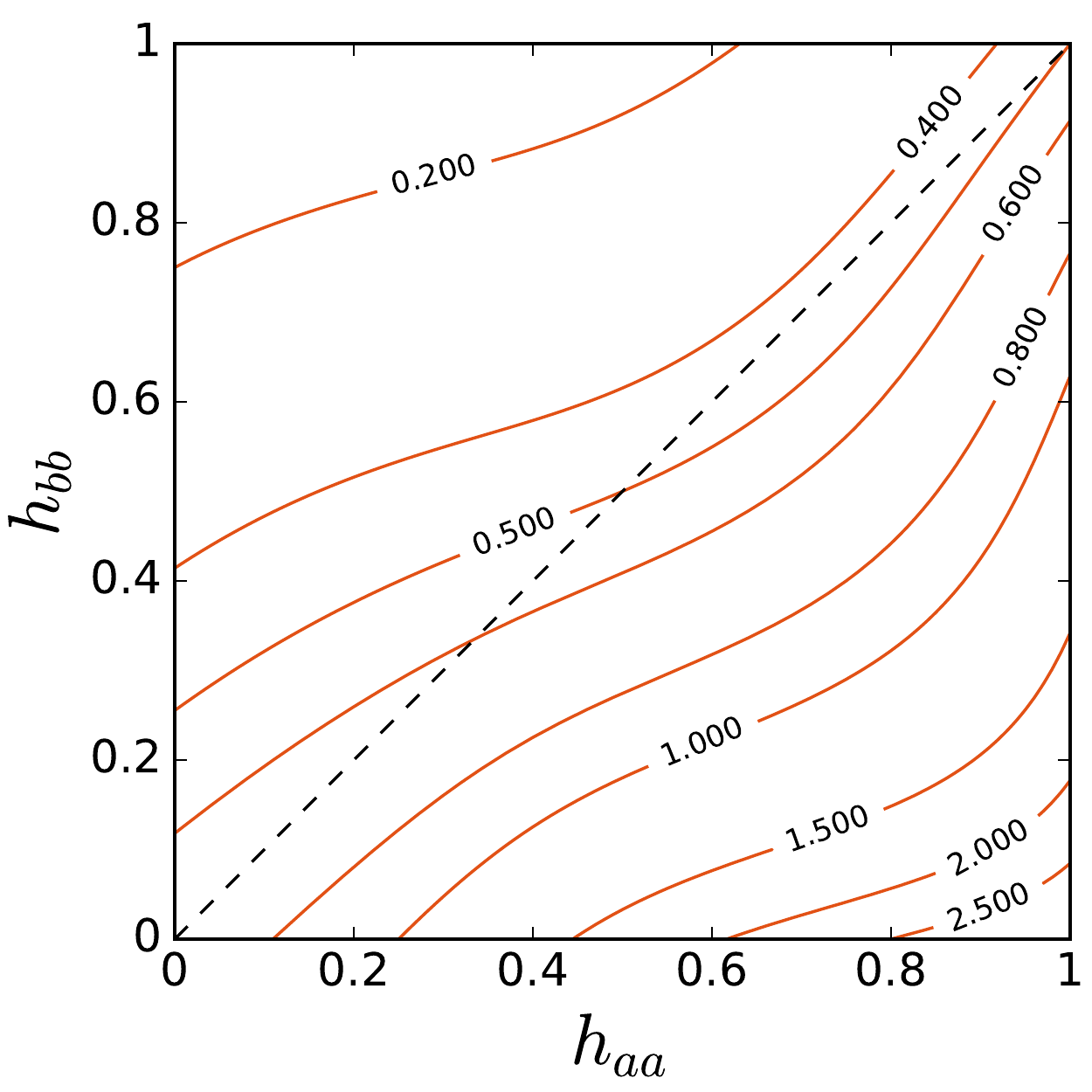}
\includegraphics[width=0.23\linewidth]{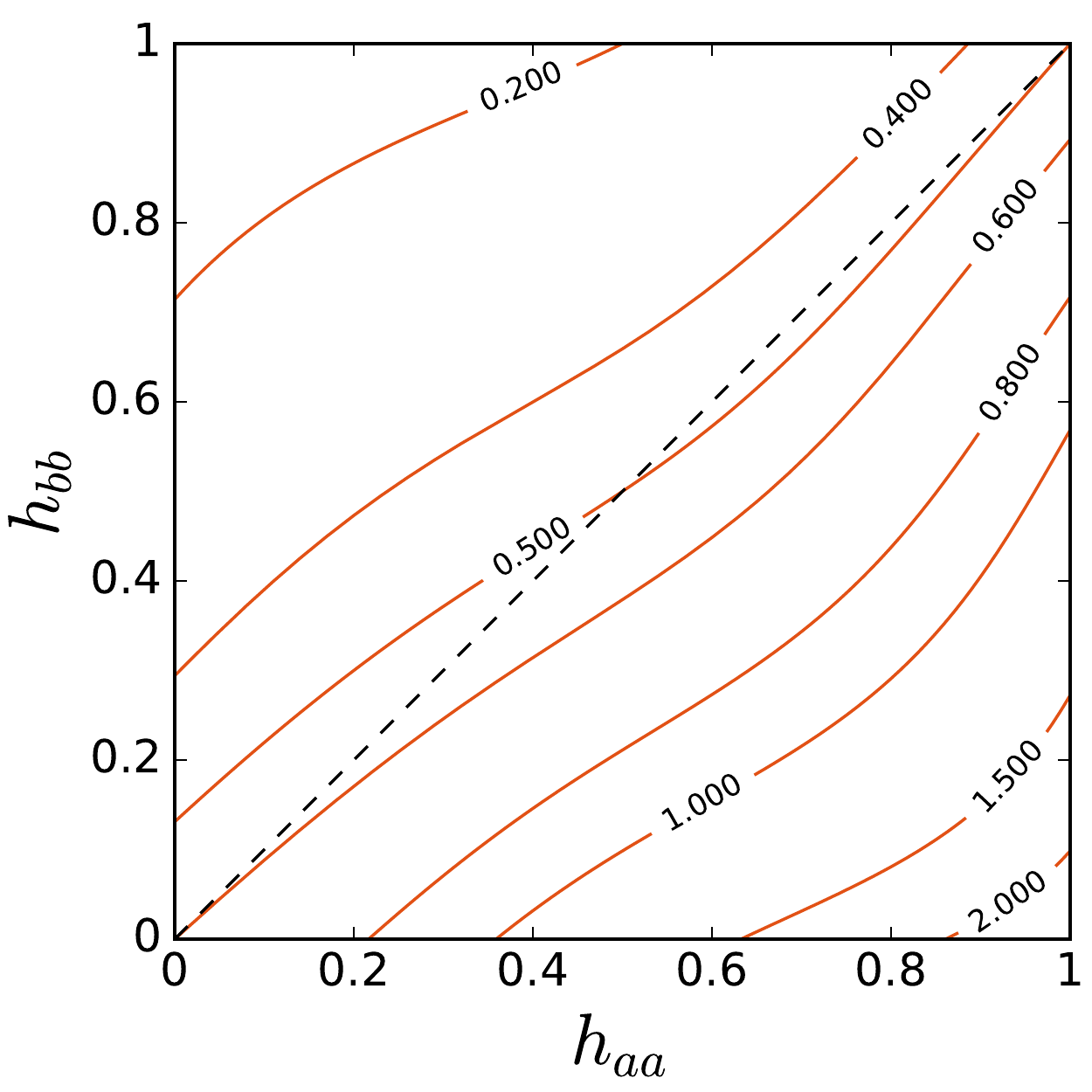}
\caption{\textbf{Evolution of the exponents for the degree growth, asymmetrical homophily.} The exponents $\beta_a$ and $\beta_b$ are defined in eqs.~(\ref{eq:ca}) and (\ref{eq:cb}). $h_{aa}$ and $h_{bb}$ are the homophily parameters. Bottom row shows the behaviour of $\beta_a$ and top row the behaviour of $\beta_b$. Columns are ordered according to the fraction of nodes belonging to the majority group (parameter $f_b$), respectively $f_a = 0.1$, 0.2, 0.3 and 0.4 from left to right. The dashed red lines indicate the symmetrical case plotted in Fig.~\ref{fig:exponents}.}
\label{fig:exponents_asym}
\end{figure*}

This equation for $C$ can be numerically solved. Within the ranges of values of the parameters, it has three real solutions, but only one in the interval $[0,2]$ and thus valid in this case. We can then derive the evolution of the degree of a single node for both groups in the general case. Let's define:
\small
\begin{equation}
\begin{split}
  Y_a(t) &= h_{aa}K_a(t) + h_{ab}K_b(t)\\
  &= h_{aa}Cmt + h_{ab}(2-C)mt\\
  &= mt(h_{aa}C + h_{ab}(2-C))\\
\end{split}
\end{equation}
and
\begin{equation}
\begin{split}
  Y_b(t) &= h_{ba}K_a(t) + h_{bb}K_b(t)\\
  &= h_{ba}Cmt + h_{bb}(2-C)mt\\
  &= mt(h_{ba}C + h_{bb}(2-C))\\
\end{split}
\end{equation}
\normalsize
For group $a$, we have:

\begin{equation}\label{eq:ca}
\begin{split}
  \dfrac{dk_a}{dt} &= mf_a\dfrac{h_{aa}k_a}{Y_a} + mf_b\dfrac{h_{ba}k_a}{Y_b}\\
  &= \dfrac{k_a}{t}\left(\dfrac{f_ah_{aa}}{h_{aa}C + h_{ab}(2-C)} + \dfrac{f_bh_{ba}}{h_{ba}C + h_{bb}(2-C)}\right)\\
  &= \dfrac{k_a}{t}\beta_a\\
\end{split}
\end{equation}
and thus:
\begin{equation}
  k_a(t) \propto t^{\beta_a}\\
\end{equation}
Similarly, for group $b$ we have:
\begin{equation}\label{eq:cb}
\begin{split}
  \dfrac{dk_b}{dt} &= mf_b\dfrac{h_{bb}k_b}{Y_b} + mf_a\dfrac{h_{ab}k_b}{Y_a}\\
  &= \dfrac{k_b}{t}\left(\dfrac{f_bh_{bb}}{h_{ba}C + h_{bb}(2-C)} + \dfrac{f_ah_{ab}}{h_{aa}C + h_{ab}(2-C)}\right)\\
  &= \dfrac{k_b}{t}\beta_b\\
\end{split}
\end{equation}
and thus:
\begin{equation}
  k_b(t) \propto t^{\beta_b}
\end{equation}

We plot the evolution of these exponents $\beta_a$ and $\beta_b$ in the special case where $h_{aa} = h_{bb} = h$ and $h_{ab} = h_{ba} = 1-h$ (Fig.~\ref{fig:exponents}). The general case where homophily is not symmetrical is shown in the contour plot in figure \ref{fig:exponents_asym}. The dashed red lines indicate the previous case of symmetric homophily.


Finally, as has been shown before, there is an inverse relation between the exponent of the degree growth and the exponent of the degree distribution ($p(k) \propto k^\gamma$), as follow \cite{Barabasi99,bianconi2001competition} :

\begin{equation}
\label{eq:degree_exponent_SI}
    \gamma =\frac{1}{\beta} + 1
\end{equation}

In the case where homophily is equal to 0.5 for both groups, we have  $\beta_{a} = \beta_{b} = 0.5$, in which the model converges to classic BA model with degree exponent $p(k) \propto k^{-3}$.  

\subsection{Estimating asymmetric homophily parameters}

\begin{figure*}[]
\centering
\includegraphics[width=0.255\linewidth]{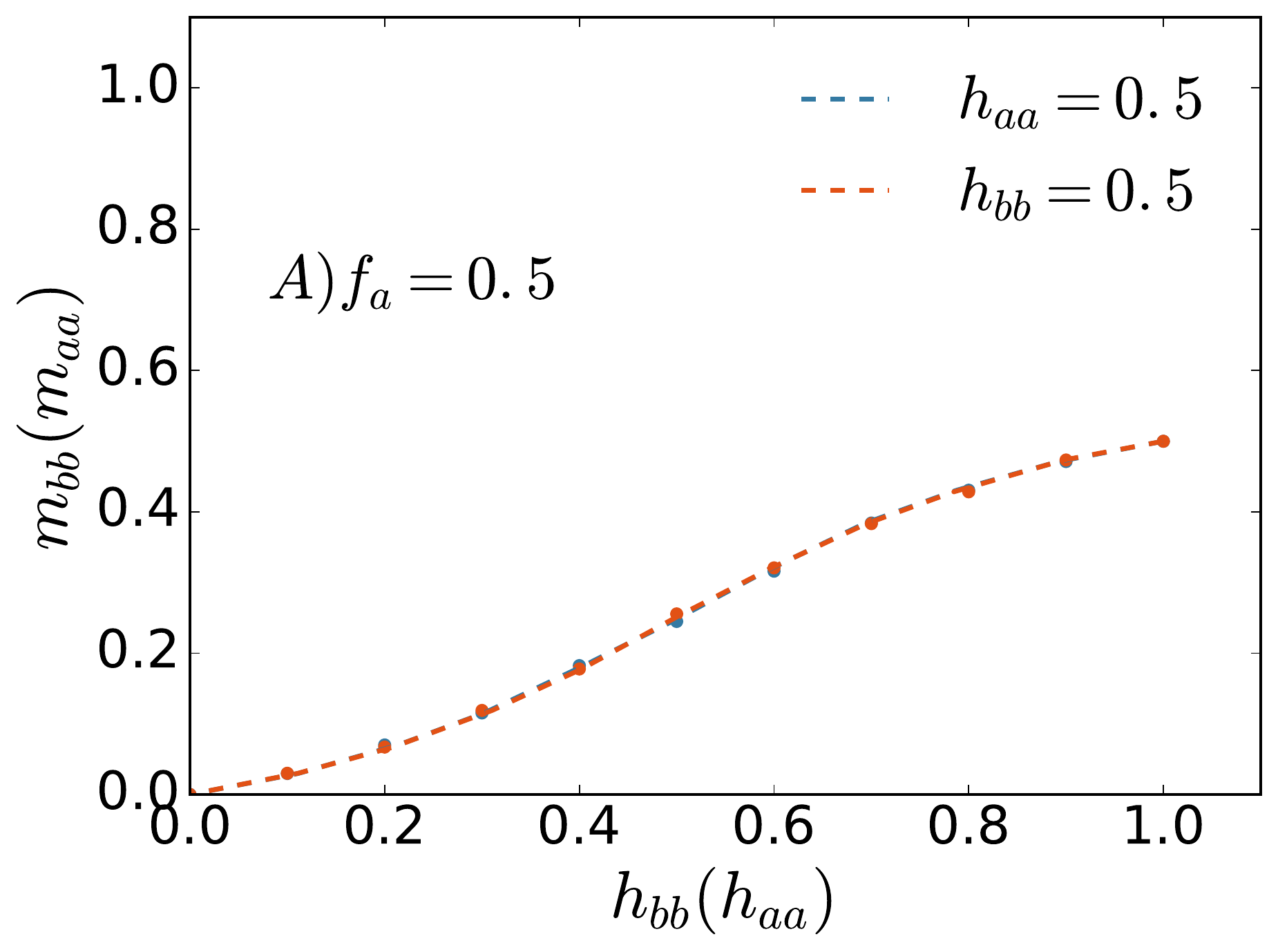}
\includegraphics[width=0.24\linewidth]{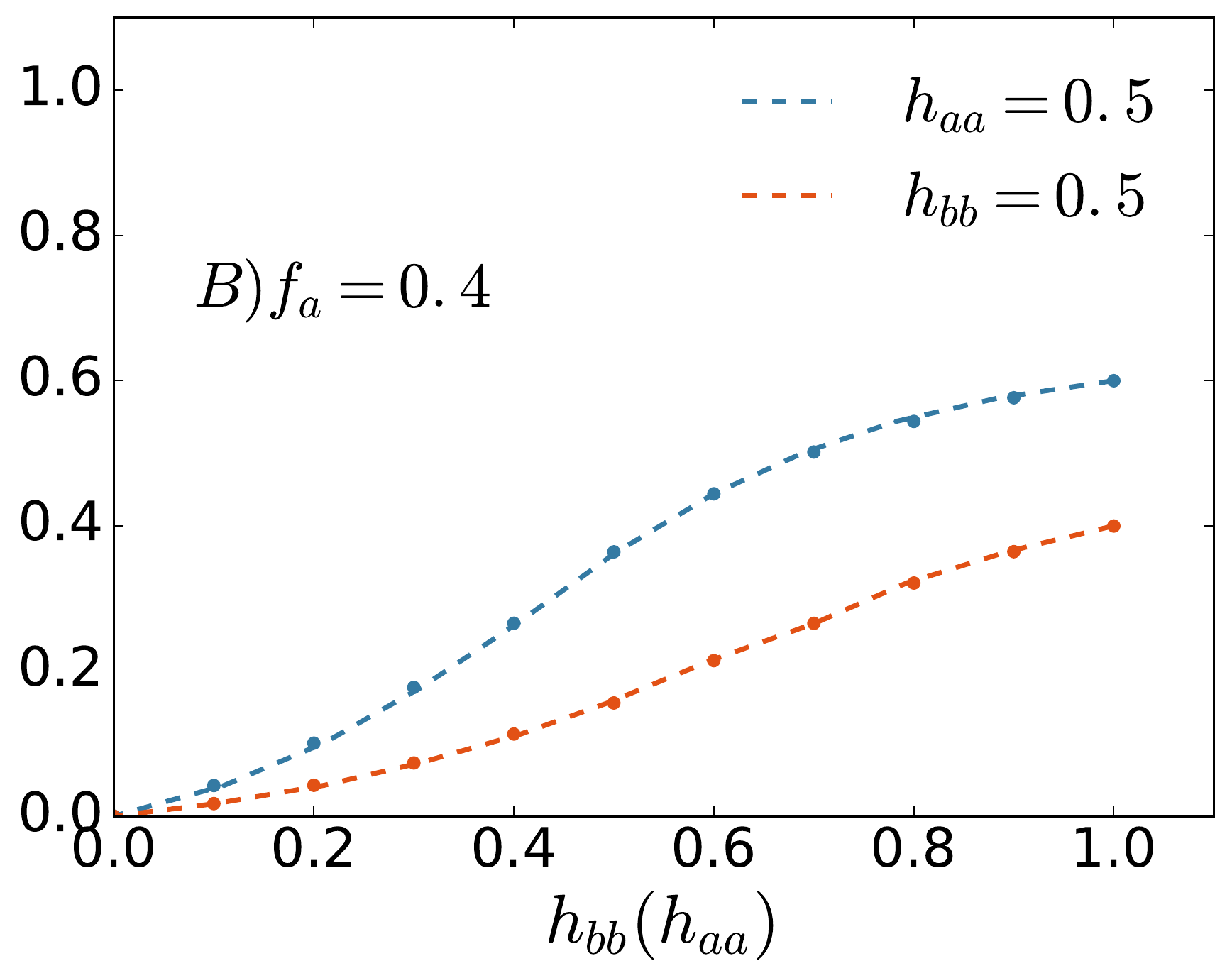}
\includegraphics[width=0.24\linewidth]{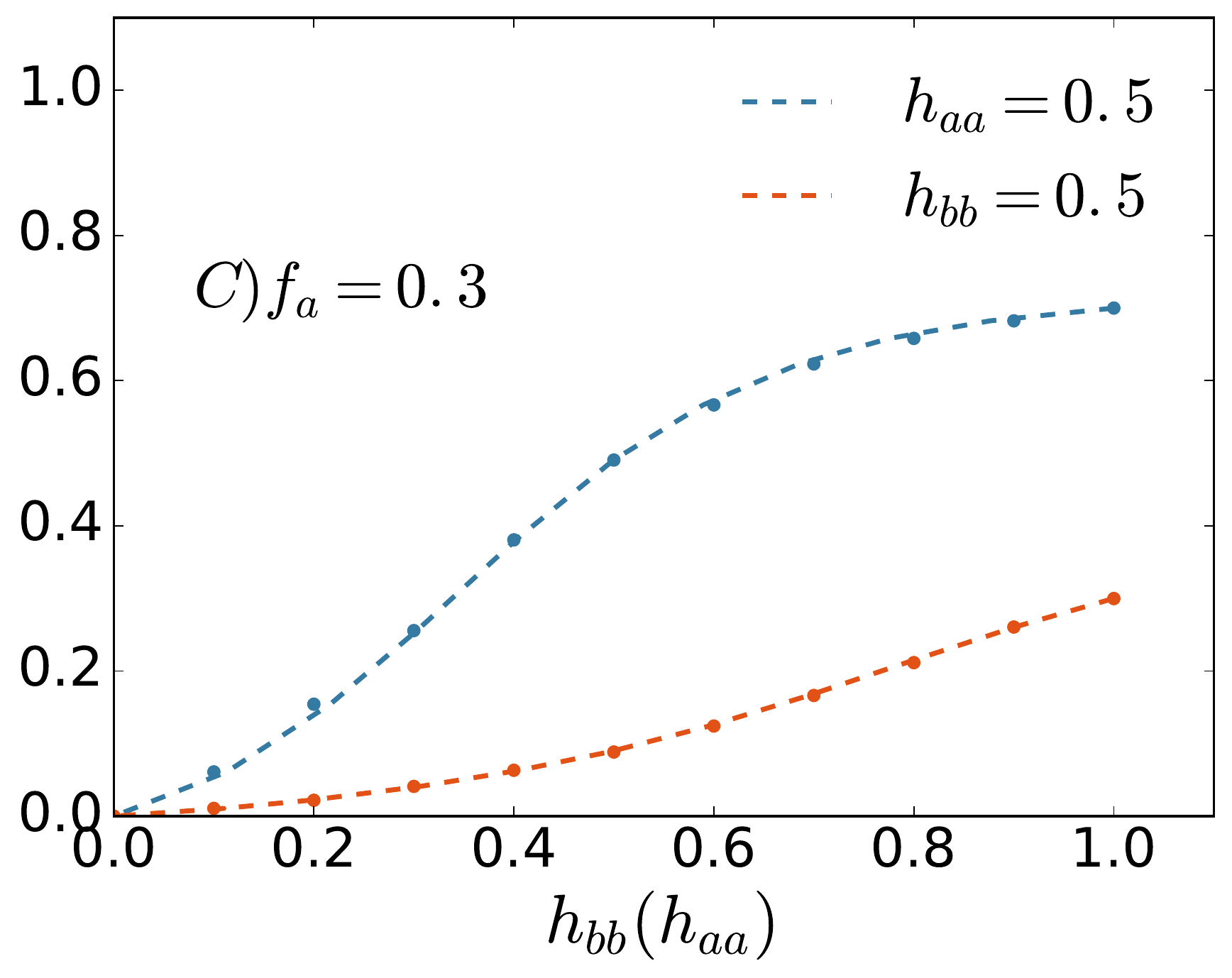}
\includegraphics[width=0.24\linewidth]{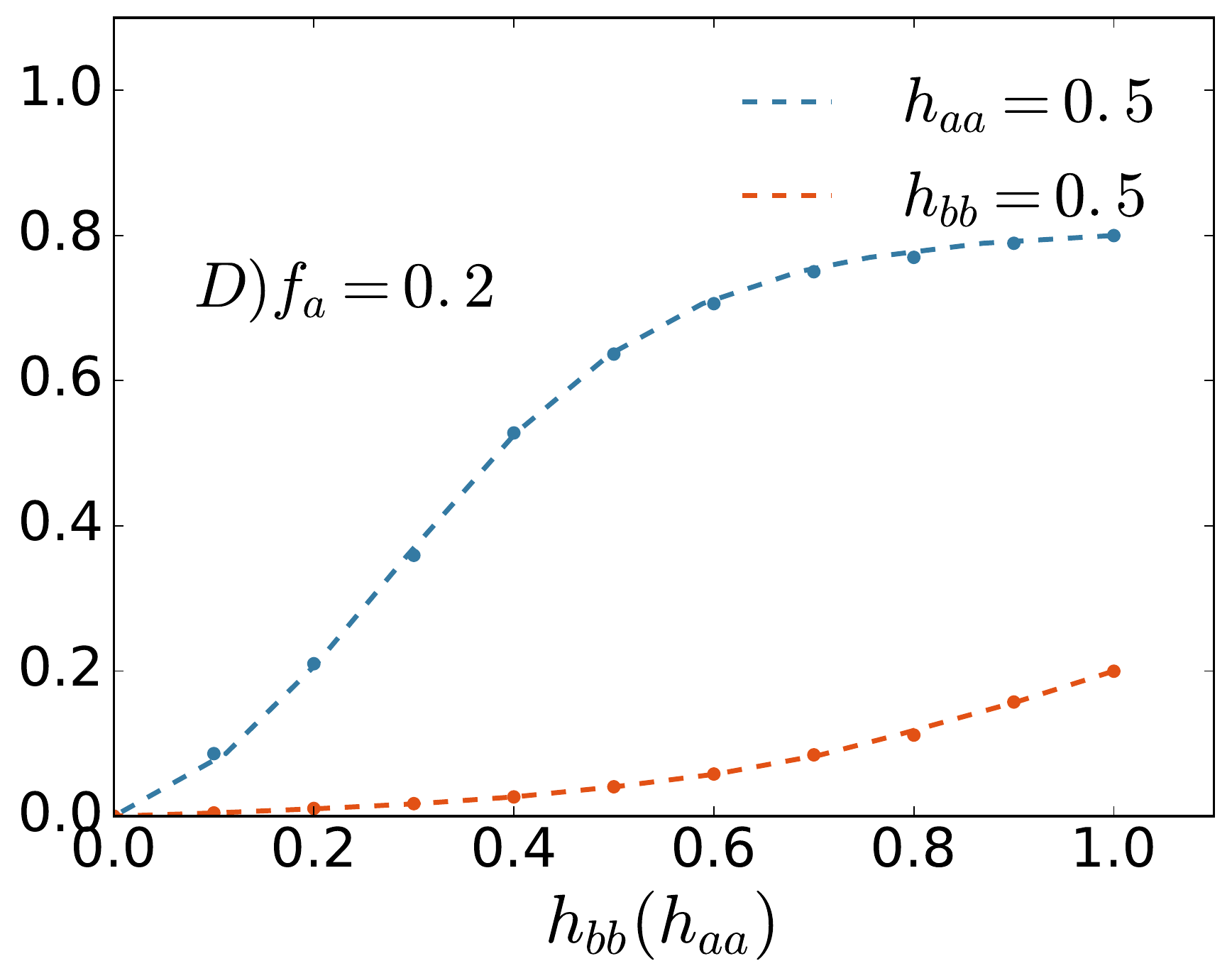}

\caption{\textbf{Analytical and numerical estimation of the fraction of edges that run within each group of nodes versus homophily.} Fractions of edges within each group are denoted by $m_{aa}$ and $m_{bb}$. The homophily parameter is tuned for one group and fixed for another group.  Panels from left to right are generated for various minority sizes. The numerical results are shown by points in the plot. The analytical results are shown by dashed lines. As the size of the minority decreases, the gap between the fraction of edges for minority (orange lines when majority homophily is fixed ($h_{bb} = 0.5$)) and majority (blue lines when minority homophily is fixed ($h_{aa} = 0.5$)) widen.
The analytical results are derived by estimating expected homophily from number of edges and they are in excellent agreement with the numerical results. }
\label{fig:analytical_edges}
\end{figure*}

The analytical derivations in the previous section enable us to estimate the homophily parameter given the fraction of edges that exist within each group in empirical networks. 


In a network with $M$ number of edges, let's assume $M_{aa}$ is the number of edges linking two nodes of the group $a$  (ingroup links) and similarly $M_{bb}$ is the number of edges linking nodes of the group $b$. The probability to have an ingroup link in group $a$ can then be defined as $m_{aa} = \frac{M_{aa}}{M}$, which depends on the group size $f_a$, the homophily parameter $h_{aa}$ and the relative degree growth exponents $\beta_{a}$ and $\beta_{b}$:
\begin{equation}
m_{aa} = \frac{f_{a}^2 h_{aa}\alpha_{a}}{f_{a}^2 h_{aa}\alpha_{a} + f_{a} f_{b} h_{ab}\alpha_{b}}
\end{equation}

A similar formula can be written for the group $b$:
\begin{equation}
m_{bb} = \frac{f_{b}^2 h_{bb}\alpha_{b}}{f_{b}^2 h_{bb}\alpha_{b} + f_{b} f_{a} h_{ba}\alpha_{a}}
\end{equation}

where $\alpha_{a} = \frac{\beta_{a}}{\beta_{a} + \beta_{b}}$ and $\alpha_{b} = \frac{\beta_{b}}{\beta_{a} + \beta_{b}}$ are relative degree exponents for each group. 

Note that in the general case homophily can be asymmetric, $h_{ab} \neq h_{ba}$. From our previous analytical calculations, we know the relation between the exponent $\beta$, the group size and the homophily parameter by numerically solving equation \ref{eq:C} given $m_{aa}$ and $m_{bb}$. We can then solve these nonlinear dynamical equations and determine the expected homophily $h_{aa}$ and $h_{bb}$ for group $a$ and $b$. 

Results are shown in figure~\ref{fig:analytical_edges}. For simplicity, we fix the value of the homophily parameter in one group and show the relation between tunable homophily and the fraction of edges for the other group. The dashed lines corresponds to the results of the analytical derivation, given a number of edges for each group. The value of the homophily parameter extracted from the simulations is shown by the dots.
In the case of homophily fixed for one group at $0.5$ and same group size (panel left), we observe as expected a sigmoid function for both groups. For large value of homophily ($h_{aa}, h_{bb} = 1$), the fraction of edges between nodes of the same group converges to the size of the group. As the size of the minority decreases, the gap between the fraction of edges for the minority (orange lines when the majority homophily is fixed ($h_{bb} = 0.5$)) and the majority (blue lines when the minority homophily is fixed ($h_{aa} = 0.5$)) widen. By tuning the group size and fixing the homophily parameter for minorities, the majority gains an advantage by receiving links within itself partly because of the increase in their degree exponent and large group size differences (blue lines). 



\bibliography{ref}



\section*{\large Acknowledgements}
We thank Daniele Cassese and Renaud Lambiotte for their invaluable comments on the paper.
\section*{\large Additional information}
Code and data can be found in Github repository \url{https://github.com/frbkrm/HomophilicNtwMinorities.git}

\smallskip

\begin{appendices}
\cleardoublepage
\let\clearpage\relax
\noindent{\huge\bfseries APPENDIX\par}
%
%

\section{Degree distribution - continuum approximation}



Here we use continuum theory similar to \cite{bianconi2001competition} to compute the degree growth for each group. The probability that a newly entered node $j$ choose node $i$ depends on the relative ``fitness'' of the node $i$ towards the node $j$. We approximate the relative fitness by averaging fitness depends on the probability of visiting each group of nodes. Let us denote the average fitness of a node in group $a$ as $\bar{h_{a}}$.

The rate of degree growth for a node in each time step depends on number of arrival links $m$, the relative fitness of the node to other nodes and its degree:

\begin{equation}
\label{eq:k_growth}
    \frac{\partial k_i}{\partial t} = m \frac{ \bar{h_{i}} k_i}{\sum_{l} \bar{h_{l}} k_l}
\end{equation}
 
At each time step, node $i$ has $m$ chances to be selected. The sum in the denominator goes over all links that occur from $t=0$ to $t = t$.

Assuming that the node $i$ joins the network at time $t_0$, the average degree of the node at time $t$ is
\begin{equation}
\label{eq:degree_growth}
    k_i(t,t_{0}) = m  (\frac{t}{t_0})^{\beta(\bar{h_{i}})}
\end{equation}

where we define
\begin{equation}
\label{eq:beta}
    \beta(\bar{h_{i}}) = \frac{ \bar{h_{i}} }{\sum_{l} \bar{h_{l}} k_l} 
\end{equation}


we can rewrite Eq.~\ref{eq:k_growth} as follows:
\begin{equation}
\label{eq:k_growth2}
\frac{\partial k_i}{\partial t} = m k_i \beta(\bar{h_{i}})
\end{equation}

The exponent $\beta(\bar{h_{i}})$ is bounded, $0< \beta(\bar{h_{i}})< 1$, because number of links always increase for a node and number of links cannot increase faster than $t$. 

The main difference of this model with the classical preferential attachment model is the parameter $\bar{h_{i}}$ which regulates the homophily between node $i$ to all other nodes. If we assume that node $i$ belong to group $a$, then the average fitness of nodes in group $a$, $\bar{h_{a}}$, depends on probability of visiting nodes from the same group multiply their relative homophily and probability of visiting nodes from different group and their relative homophily:

\begin{equation}
\label{eq:fitness_group}
    \bar{h_{a}} = f_{a} \delta(j-a) (h_{ja}) + \delta(j-b)  f_{b} (h_{jb})
\end{equation}

In which $f_{a}$ and $f_{b}$ represent the fraction of nodes in group $a$ and $b$ and $h_{ja}$ relative homophily between node from group $j$ to group $a$.


Let's now consider the mean of the sum in the denominator of the exponent in equation \ref{eq:beta}. In the continuum approximation, the sum can be written as integral over nodes that are born in different time ($t_{0}$):

\begin{equation}
   \ev{\sum_{l} h_{l} k_l} = \int \bar{h_{l}} d\bar{h} \rho(\bar{h}) \int_{1}^{t} dt_{0} k_i(t,t_{0}) 
\end{equation}

In the case of homophilic graph with two groups, $\rho(\bar{h})$ is equivalent to the size of each group.

Inserting $k_i(t,t_{0})$  into the equation, and neglecting $t^{\beta}$ when $t$ is large, $t \to \infty$, we get,
\begin{equation}
   \ev{\sum_{l} h_{l} k_l} = C m t
\end{equation}
where 
\begin{equation}
\label{eq:c}
   C = \int \rho(\bar{h}) d\bar{h} \frac{h_{ij}}{1-\beta(\bar{h})}
\end{equation}

$C$ is the denominator of eq.~\ref{eq:beta} and it shows the growth of connectivity probability. Inserting that into Eq.~\ref{eq:k_growth2}, we get
\begin{equation}
\label{eq:beta_2}
    \beta(\bar{h}) = \frac{\bar{h}}{C}
\end{equation}

Note that in the presence of two groups, $C$ consists of two parts, $C_{a}$ and $C_{b}$:
\begin{equation}
\label{eq:c_all}
   C = C_{a} + C_{b}
\end{equation}

Using eq.~\ref{eq:c} we can compute the probability growth for each group. For group $a$ we have:

\begin{align*}
\label{eq:c2}
  C_{a} &= \int_{\bar{h}} \rho(\bar{h}) d\bar{h} \frac{1}{1 - \beta(\bar{h}) } \\
    &= \frac{f_{a} h_{aa}}{1 - \beta(\bar{h_a})   } + \frac{f_{b} h_{ab}}{1 - \beta(\bar{h_b}) } 
\end{align*}

and
\begin{equation}
\label{eq:cc}
  C_{b} = \frac{f_{b} h_{bb}}{1 - \beta(\bar{h_b})   } + \frac{f_{a} h_{ba}}{1 - \beta(\bar{h_a}) } 
\end{equation}

Since in each time step only one node is arriving, the sum of all degrees should be equivalent to all the incoming nodes and links 

\begin{equation}
   \ev{\sum_{i}  k_i} =  \int_{1}^{t} dt_{0} k_i(t,t_{0}) = mt \sum_s f_s / (1 - \bar{q_s}) = 2mt
\end{equation}

therefore get an additional identity
\begin{equation}
   \sum_s f_s / (1 - \bar{{h}_s}) = 2
\end{equation}

where $s$ represents number of groups which in our case group $a$ and $b$. We then can solve the self-consistent equation and get the value of $C$ and therefore determine the exponent $\beta$ for each group. In \cite{ferretti2012features} the authors derived the generalized form of this model using rate equation approach \cite{krapivsky2000connectivity} for nodes that are distributed in geometrical space.

The cumulative probability that a node with fitness $\bar{h}$ has a degree larger than $k$ is

\begin{align*}
    P(k(t)> k) &= P(t_{0} < t(m/k)^{1/\beta}) = t (m/k) ^{C/\bar{h}}
\end{align*}

Thus the probability of a node to have $k$ links, is given by

\begin{equation}
\label{eq:degree_exponents}
    p(k) = \int_{\bar{h}} \rho(\bar{h}) d\bar{h} \frac{\partial P(k)}{\partial k} =  \int_{\bar{h}} \rho(\bar{h}) d\bar{h} \frac{C}{\bar{h}} (m/k)^{\frac{1}{\beta(\bar{h})} + 1}
\end{equation}

\begin{equation}
\label{eq:degree_exponent}
    \gamma(h) =\frac{1}{\beta(\bar{h})} + 1
\end{equation}

The slope of the distribution is determined by the exponent $\frac{C}{\bar{h}} + 1$. In the case of original Barab\'{a}si-Albert model with $C=2$ and $q = 1$, we get $p(k) \propto k^{-3}$. The same is true in our model with equal homophily for minorities and majorities. If homophily is equal to $0.5$, the attractiveness for both population is the same and the sum of probabilities over time $C(t)$ converges to $1$. Therefore the slope of the distributing will be 3.

\end{appendices}

\end{document}